\documentclass[12pt]{article}
\usepackage{graphicx,bm,amsmath,amssymb,latexsym,psfrag,epsfig,subfigure,euscript,multirow,color,verbatim}
\usepackage{bbm}

\allowdisplaybreaks

\newcommand{\ket}[1]{\left| #1 \right>} 
\newcommand{\bra}[1]{\left< #1 \right|} 

\def\tmmathbf{\mathbf}

\def\rnO{ {\red{n_1}} }

\def\nbO{ {\red{{\bar n}_1}}}

\def\nslashL{\rlap{\hspace{0.02cm}/}{ {\red n_\rF}}}
\def\nslashO{\rlap{\hspace{0.02cm}/}{ {\red n_\rO}}}

\def\la{\left|}
\def\ra{\right|}

\newcommand{\nn}{\nonumber}
\newcommand{\ds}{\displaystyle}

\def\rd{\mathrm{d}}

\def\cdt{ \hspace{-0.1em} \cdot \hspace{-0.1em} }

\def\xx{\xi}

\def\rd{\mathrm{d}}

\definecolor{darkred}{rgb}{0.7,0.0,0.0}
\definecolor{darkblue}{rgb}{0.0,0.0,0.9}
\definecolor{darkgreen}{rgb}{0.0,0.5,0.0}
\definecolor{brown}{rgb}{0.0,0.0,0.0}
\newcommand{\red}{\color{darkred}}
\newcommand{\blue}{\color{darkblue}}
\newcommand{\green}{\color{darkgreen}}

\newcommand{\bI}{ {\blue I} }
\newcommand{\bJ}{ {\blue J} }
\newcommand{\bK}{ {\blue K} }
\newcommand{\bL}{ {\blue L} }
\newcommand{\ba}{ {\blue a} }
\newcommand{\bb}{ {\blue b} }
\newcommand{\bc}{ {\blue c} }
\newcommand{\bd}{ {\blue d} }
\newcommand{\bi}{ {\blue i} }
\newcommand{\bj}{ {\blue j} }
\newcommand{\MIJ}{  M_{\bI \bJ} }

\newcommand{\GHIJ} {\Gamma^H_{\bI \bJ}}

\newcommand{\bO}{ {{\blue 1}} }
\newcommand{\bT}{ {{\blue 2}} }
\newcommand{\bTH}{ {{\blue 3}} }
\newcommand{\bF}{ {{\blue 4}} }
\newcommand{\bE}{ {{\blue 8}} }
\newcommand{\bN}{ {{\blue 9}} }
\newcommand{\bpm}{ {{\blue \pm}} }

\newcommand{\bl}{ {\blue{\lambda} }}

\newcommand{\gG}{ {\green{\Gamma} }}

\newcommand{\gpm}{ {\green{\pm} }} 
\newcommand{\gmp}{ {\green{\mp} }} 
\newcommand{\gp}{ {\green{+} }} 
\newcommand{\gm}{ {\green{-} }} 
\newcommand{\gR}{ {\green{R} }} 
\newcommand{\gL}{ {\green{L} }}

\newcommand{\gl}{ {\green{\lambda} }} 
\newcommand{\gGG}{ {\green{\Gamma}}}

\newcommand{\cC}{ {\mathcal{C} }} 
\newcommand{\cJ}{ {\mathcal{J} }}

\newcommand{\cM}{ {\mathcal{M} }} 
\newcommand{\cW}{ {\mathcal{W} }} 
\newcommand{\cO}{ {\mathcal{O} }}
\newcommand{\gcusp} {\gamma_{\mathrm{cusp}}}

\newcommand{\rF}{ {{\red 4}} }
\newcommand{\rTH}{ {{\red 3}} }
\newcommand{\rO}{ {{\red 1}} }
\newcommand{\rT}{ {{\red 2}} }

\newcommand{\wh}{\widehat}
\newcommand{\what}{{}}

\begin{document}
\begin{titlepage}

\begin{flushright}
\end{flushright}

\vspace{0.2cm}
\begin{center}
\Large\bf
1-loop matching and NNLL resummation\\
for all partonic  2$\to$2 processes in QCD
\end{center}

\vspace{0.2cm}
\begin{center}
{\sc Randall Kelley and Matthew D. Schwartz}\\
\vspace{0.4cm}
{\sl Center for the Fundamental Laws of Nature \\
Harvard University\\
Cambridge, MA 02138, USA}
\end{center}

\vspace{0.2cm}
\begin{abstract}\vspace{0.2cm}
\noindent 
The Wilson Coefficients for all 4-parton operators which arise in matching QCD to  Soft-Collinear Effective Theory (SCET)
are computed at 1-loop. 
Any dijet observable calculated in SCET beyond leading order will require these
results. The Wilson coefficients are separated by spin and color, although most applications
will involve only the spin-averaged hard functions.
The anomalous dimensions for the  Wilson coefficients are given to 2-loop order, and the renormalization
group equations are solved explicitly. This will allow for analytical resummation
of dijet observables to next-to-next-to-leading logarithmic accuracy.  For each channel, there is a natural
basis in which the evolution is diagonal in color space. 
The same basis also diagonalizes the color evolution for the soft function. Even though soft functions
required for SCET calculations are observable dependent, it is shown that their renormalization group
evolution is almost completely determined by a universal structure.
With these results, it will be possible to calculate hadronic event shapes
or other dijet observables to next-to-leading order with next-to-next-to-leading log resummation. 
\end{abstract}
\vfil
\end{titlepage}

\section{Introduction}
While the main goal of the Large Hadron Collider at CERN is to find evidence for physics
beyond the standard model, the standard model itself must be accurately understood
before any new physics claim can be made. The most common way of comparing
data to predictions of the standard model is through Monte Carlo simulations. While
these simulations are remarkably accurate in reproducing the gross kinematic
features (and often the fine structure) of observables, 
they are formally accurate to only leading-order in
perturbation theory and with resummation of only the leading Sudakov
double-logarithmic singularities (LL). While much theoretical work has been
devoted to computing phenomenologically relevant observables to next-to-leading
order (and sometimes even next-to-next-to-leading order), very little is
known about the effects of higher-order resummation. Since observables
at the LHC generally involve many scales, it is likely that resummation
will lead to a quantitatively significant improvement of our theoretical predictions.
Recent advances in Soft-Collinear Effective Theory have shown resummation
to be achievable at the 
next-to-leading (NLL)~\cite{Hornig:2009vb,Schwartz:2007ib}, 
next-to-next-to-leading (NNLL)~\cite{Manohar:2003vb,Idilbi:2005ky,Becher:2009th,Ahrens:2010zv} and
next-to-next-to-next-to-leading (N${}^3$LL) 
logarithmic order
~\cite{Idilbi:2005ni,Idilbi:2006dg,Becher:2007ty,Becher:2008cf,Ahrens:2008nc,Chien:2010kc,Abbate:2010xh}.
Practical results have included a precise measurement of $\alpha_s$ from event shapes~\cite{Becher:2008cf,Ahrens:2008nc},
better agreement with Tevatron data and reduced theoretical uncertainty for the the direct photon $p_T$ spectrum~\cite{Becher:2009th}, 
and a bound on possible new colored states~\cite{Kaplan:2008pt}. In all of these cases, higher order resummation was
critical to the improved theoretical precision, and so it is natural to ask whether resummation might be similarly
important for pure QCD events.

In this paper, we take the first step towards the calculation of dijet observables
at next-to-leading order (NLO) with NNLL resummation. 
The computation of a dijet observable in SCET requires a factorization formula which convolutes 
parton distribution functions (PDFs), Wilson coefficients, jet functions and a soft function. Although
the jet and soft functions are observable dependent, the Wilson coefficients are universal and independent
of the details of the factorization theorem. For an NLO calculation, the
complete set of NLO Wilson coefficients for all the $2\to 2$ processes of QCD is required,
but explicit expressions
have only been presented in the literature for 
the
$q q \to q q$ channel and its various crossings~\cite{Chiu:2008vv}. 
 The necessary diagrams 
for $q \bar{q} \to gg$  and its crossings were computed in~\cite{Fuhrer:2010eu,Kunszt:1993sd}
and for $gg\to gg$ in~\cite{Kunszt:1993sd} and~\cite{Bern:1990cu}. These amplitudes 
can be used to extract the Wilson coefficients for the remaining channels. This computation,
and the compilation and simplification of all of the 1-loop expressions, is a main result of this paper.

The other main result of this paper is the elucidation of how the color structures and spin states 
interfere in the context of resummation.
Processes with four hard partons in QCD separate into a number of color structures, corresponding
to separate operators in SCET, which mix under RG evolution. Similar color mixing 
has already been studied for collinearly-regulated soft Wilson lines  
using the traditional approach to resummation~\cite{Kidonakis:1998nf}, and for quark scattering~\cite{Chiu:2008vv},
gauge boson production~\cite{Chiu:2009mg} and $t\bar{t}$ production~\cite{Ahrens:2010zv} using SCET.
For NNLL resummation, the renormalization group equations must be solved to least 2-loop order.
A general form for these RGEs
is known~\cite{Chiu:2009mg,Becher:2009cu,Becher:2009qa,Gardi:2009qi,Dixon:2009ur,Dixon:2010hy}, 
as is the related soft function evolution equation~\cite{Kidonakis:1998nf}.

Since jet functions and PDFs are color diagonal, the color
mixing in the  evolution of the Wilson coefficients is exactly compensated for by color mixing in the soft
function evolution. 
In fact, since the Wilson coefficients are universal (observable independent),
the mixing terms in the soft function must also be universal.
Moreover, all of the color-mixing is proportional to the universal cusp anomalous dimension 
(up to at least 2-loop order). Therefore, the color-mixing effects can be diagonalized 
once-and-for-all by a careful  choice of operator basis. 
This natural basis for evolution is different from the one that 
is natural for matching, in which the Wilson coefficients take a particularly simple form. 
This implies that although the color mixing in the evolution
can be diagonalized, there are still additional interference effects relevant beginning
at NNLL due to the non-diagonal components of the NLO hard and soft functions.

This paper is organized as follows. In Section~\ref{sec:general}, we give an overview of factorization for dijet processes
in SCET. Section~\ref{sec:operators} introduces the operators for the $2\to 2$ processes. 
Their Wilson coefficients are computed in Section~\ref{sec:wc} and the RGEs are presented 
and solved in Section~\ref{sec:RGE}. Then, in Section~\ref{sec:soft} we explain which features 
of the soft function evolution are universal. Finally,~\ref{sec:conc} presents
some conclusions and possible applications of these results.

\section{Factorization in SCET\label{sec:general}}
Cross sections for dijet observables calculated with Soft-Collinear
Effective Theory~ \cite{Bauer:2000yr,Bauer:2001yt,Beneke:2002ph} will have the general form
\begin{equation} \label{genform}
 \rd \sigma \sim \rd \Pi
 \sum_{\overset{\bI, \bJ, \gG,}{\mathrm{channels}}}\frac{1}{N_{\text{init}}}
\cC_{{\bJ}}^{\gG \star} \cdot S_{\bJ \bI}\cdt \cC_{{\bI}}^{\gG} \otimes \cJ \otimes \cJ 
 \otimes f \otimes f \,.
\end{equation}
Here, $\rd\Pi$ is the Lorentz-invariant phase space and $N_\text{init}$ are the number of initial states, which
are averaged over in computing the cross section.
 $\cC_{\bI}^{\gG}$ are Wilson coefficients, with $\bI$ indexing the different color
structures and $\gG$ indexing the different spins. 
The Wilson coefficients encode information about the hard $2\rightarrow 2$ scattering process.
$S_{\bI \bJ}$ are the soft functions. These form
a matrix in color space, but are independent of spin. The jet functions $\cJ$ and the PDFs $f$ are spin-averaged and 
color-independent.
The sum is also over channels, which for $2 \rightarrow 2$
processes are $q q \rightarrow q q$, $\bar{q} q \rightarrow g g$
and $g g \rightarrow g g$ and their various crossings (like $q g \rightarrow q
g$). Everything has an implicit channel index, which we suppress for clarity.
One can show that at leading power in the SCET expansion the channels do not interfere and the cross section for each
channel can simply be added together. For more details, see for example~\cite{Becher:2009th}.

The effective field theory calculation begins by matching from QCD to SCET. This involves
enumerating the appropriate operators in both theories and calculating matrix elements in relevant final states.
Wilson coefficients for the SCET operators are then adjusted so that they reproduce the amplitudes from QCD to the desired order
in perturbation theory, if evaluated at the hard scale $\mu_h$. If the SCET operators are evaluated at lower scales,
their matrix elements change, and include the effects of resummation.
Matrix elements of outgoing collinear fields give jet functions, incoming collinear fields give PDFs, and matrix elements
of soft Wilson lines give the soft function. Due to factorization, all of these objects can be evaluated independently and combined together using a factorization formula.

To illustrate the procedure in detail, consider the
process  $q q' \rightarrow q q'$.
We will label the incoming partons as ${\red 1}$ and ${\red 2}$ and the outgoing partons as $\rTH$ and $\rF$.
The operators in QCD appropriate for this channel have the form 
\begin{equation}
 \mathcal{O}^{\text{QCD}}_{\bI\gGG } =
 ( \bar{q}_\rF {\blue T_I} \gamma_{\mu} \gG q_\rT ) 
 ( \bar{q}_\rTH {\blue T_I} \gamma^{\mu} \gG'q_\rO )  \,.
\end{equation}
Here, ${\blue I}$ indexes the color structure (${\blue T_\bO} = {\blue \tau^a}$ or ${\blue T_\bT} = {\blue \mathbf{1}}$),
and $\gG$ and $\gG'$ index the spin (e.g. $\gG =\gG'= {\green P_L} =  \frac{1}{2}\left( 1 -\gamma_5\right)$).
The operators in SCET follow from replacing each QCD quark with a collinear
quark and adding soft and collinear Wilson lines,
\begin{align} \label{qqops}
 \cO_{\bI\gGG}^{\text{SCET}} &=
( \bar{\chi}_\rF Y_\rF^\dagger {\blue T_I}\gamma_{\mu}  \gG Y_\rT \chi_\rT) 
( \bar{\chi}_\rTH Y_\rTH^\dagger {\blue T_I}\gamma^{\mu}  \gG Y_\rO \chi_\rO)  \,.
\end{align}
The fields denoted by $\chi_{\red i}$ are collinear quarks in the ${\red n_i^{\mu}}$ direction wrapped in
 ${\red n_i}$-collinear Wilson lines (not shown). The $Y_{\red i}$ are soft Wilson lines in the fundamental representation of $SU(3)$.
For more details of the notation, see for example~\cite{Becher:2009th}. 

In SCET there is a separate set of collinear degrees of freedom associated with each direction 
$\rO \rT \rTH \rF$,
and a single set of soft degrees of freedom which can communicate among the collinear sectors.
When we calculate a cross section from the squared matrix elements of SCET operators,
contributions from different collinear sectors do not interfere and can be factorized into 
separate calculations.  Vacuum matrix elements of collinear fields associated with outgoing
directions turn into jet functions. For example, we may be interested in
the inclusive jet function in the $\rF$ direction. This requires the evaluation of
\begin{align}\label{eq:jetdef}
\langle  0|\, \bar \chi_\rF\left( x \right) \gG   \chi_\rF(0) \, | 0 \rangle & 
=  {\rm tr}\left[\frac{\nslashL}{2} \gG \right] 
\int \frac{\rd^4 p}{(2\pi)^3} \theta(p^0)\, ({\bar{\red n}}_\rF\cdt  p ) \,\cJ_\rF(p^2)\, e^{-i\, x\, p} \, .
\end{align}
Gluon jet functions are defined similarly~\cite{Becher:2009th}, and 
less inclusive jet functions are possible as well~\cite{Jouttenus:2009ns,Cheung:2009sg,Ellis:2010rw}.
When the collinear field is associated with an incoming direction, the matrix elements
should be taken in the state of the appropriate nucleon ($N_\rO$ or $N_\rT$), which produces parton
distribution functions (PDFs). For example, the incoming collinear field in the $\rO$ direction gives
\begin{align}\label{eq:PDFdef}
\langle  N_\rO | \bar \chi_\rO (x_+) 
 \gG   \chi_\rO(0) \,|  N_\rO \rangle &=\frac{1}{4}
\nbO \cdt P_\rO \, {\rm tr}\left[ \nslashO \gG \right] \int_{-1}^1 \rd\xx\, f_{q/N_\rO}(\xx)\,e^{i\,\xx\, (\rnO \cdot x)(\nbO \cdot P_\rO)/2} \,.
\end{align}
Here, $f_{q/N_\rO}(\xx)$ is a PDF. 
The dependence on $x_+ = \rnO \cdt x \frac{ {\red \bar{n}_1^\mu} }{2}$ indicates that,
in contrast to the jet functions, fluctuations of initial states in the transverse direction are power suppressed.

The final contribution to a cross section from the SCET operators comes from the soft Wilson lines. For the operator
in Eq.~\eqref{qqops}, these are
\begin{align}
 \label{wline}
 \mathcal{W}_{\bI} 
 &= \tmmathbf{T} 
  \left\{ 
    (Y^\dagger_\rF  {\blue T_\bI} Y_\rT)^{\bi_\bF }_{\ \bi_\bT}
    (Y^\dagger_\rTH {\blue T_\bI} Y_\rO)^{\bi_\bTH}_{\ \bi_\bO}  
  \right\} 
 \nn \\ 
 &= \tmmathbf{T} 
  \left\{ 
    (Y^\dagger_\rF {\blue T_\bI} Y_\rT )
    (Y^\dagger_\rTH{\blue T_\bI} Y_\rO)  
  \right\} \,. 
\end{align}
$\mathcal{W}_\bI$ has 4 fundamental color indices which have been suppressed in the second line
since the index structure is clear. Taking vacuum matrix elements of these 
Wilson lines produces a soft function
\begin{equation}
    S_{\bI \bJ} \left( \{k\}, {\red n^{\mu}_i} \right) 
    =
    \sum_{X_s}
    \bra{0   } \mathcal{W}_\bI^\dagger \ket{X_s}
    \bra{X_s } \mathcal{W}_\bJ         \ket{0}  
    F_S (
\{k\})  \,,
\end{equation}
where the sum is over soft radiation in the final state. The function $F_S(\{k\})$ 
encodes the dependence on various projections related
to the definition of the observable.  At tree level, the soft wilson lines are trivial 
($Y_{\red i} = \blue{ \mathbbm{1} }$)
and the function $F_S(\{k \})$ must reduce to a product of delta functions since there is no soft radiation.
For the example being considered ($q q' \to q q'$), we have $\cW_\bO = {\blue \tau^a }$ and
$\cW_\bT = {\blue \mathbbm{1} }$ and the tree level soft function takes the form
\begin{align}
S^\text{tree}_{\bI \bJ}(\{ k \}, {\red n_i} )
&=  
\begin{pmatrix}
  \text{Tr}[{\blue \tau^b \tau^a }] \text{Tr}[{\blue \tau^a \tau^b }] & 
  \text{Tr}[{\blue \tau^a }] \text{Tr}[{\blue \tau^a} ]               \\
  \text{Tr}[{\blue \tau^b }] \text{Tr}[{\blue \tau^b} ]               & 
  \text{Tr}[{\blue \mathbbm{1} }] \text{Tr}[{\blue \mathbbm{1} }]     
\end{pmatrix}\prod_{ \{ k \} } \delta( k )  
\nn \\ 
&=
\begin{pmatrix}
  \frac{C_A C_F}{2} & 0 \\
  0                 & C_A^2 
\end{pmatrix}\prod_{ \{ k \} } \delta( k) \,. 
\end{align}

Note that the jet functions and PDFs are spin averaged, and the soft function is
independent of spin. Thus, the spin correlations 
are determined completely at the hard scale and are stored in the Wilson coefficients.
To be explicit, the SCET Lagrangian contains a sum over all the possible spin states $\gG$,
\begin{equation}
 {\mathcal{L}} =  \sum_{\bI,\gG}\cC_{\bI}^{\gG} \cO_{\bI}^{\gG}  \,.
\end{equation}
When one calculates a matrix-element squared from this Lagrangian, the operator products
turn into jet and soft functions which have no spin dependence. So, we can write heuristically
that
\begin{equation} \label{mform}
 | {\mathcal M}|^2 = \frac{1}{N_{\text{init}}}  \sum_{\bI, \bJ, \gG}
\cC_{{\bJ}}^{\gG \star} \cdt S_{\bJ \bI}\cdt \cC_{{\bI}}^{\gG} \otimes \cJ \otimes \cJ 
 \otimes f \otimes f .
\end{equation}
This is the same as Eq.~\eqref{genform} once the phase space factor is added. 
Therefore, for a spin-summed cross section  we only need the spin-summed hard function defined by
\begin{equation}
  H_{\bI \bJ} = \sum_\gG  \cC^{\gG}_\bI  \cC^{\gG \star}_{\bJ}\,.
\end{equation}
Then,
\begin{equation}
 | {\mathcal M}|^2 =\frac{1}{N_{\text{init}}}   \sum_{\bI, \bJ}
 H_{\bI \bJ} S_{\bJ \bI}  \otimes \cJ \otimes \cJ
 \otimes f \otimes f   \,.
\end{equation}
Since the spin information is retained in the Wilson coefficients, spin correlations can be studied using SCET,
if desired, simply by summing only over the desired spins $\gG$.

As a check on the Wilson coefficients, we can compare the $2\to2$ scattering cross section in SCET
to the spin-summed cross section in full QCD. If the matching has been done correctly, these two calculations should
agree. Although the $2 \to 2$ processes are infrared divergent and not physical, the comparison can be made in $4-2\varepsilon$ dimensions
as a formal check.
Since there is no structure to 
the outgoing or incoming partons for $2\to 2$ parton-level scattering, only virtual graphs contribute. 
Thus, the jet functions, soft functions, and PDFs 
are fixed to their tree-level values, which produce delta functions of the parton momenta. Only the soft function 
provides non-trivial structure, through its color factors. Putting these observations together,
we should find
\begin{equation} \label{qcdcheck}
  \la \cM_{(2\to 2)}^{\text{QCD}}\ra^2 =\frac{1}{N_{\text{init}}} 
\sum_{\bI \bJ}  H_{\bI \bJ} S^{\text{tree}}_{\bJ \bI} = \frac{1}{N_{\text{init}}} \text{Tr}[ H S^{\text{tree}} ]  \,.
\end{equation}
We will use this equation as an important check on our calculations, comparing the matched SCET prediction to
the cross sections from Ellis and Sexton~\cite{Ellis:1985er}. 

\section{Operators \label{sec:operators}}
In this section, we present the operators relevant for dijet production.
Again, we will label the incoming partons as ${\red 1}$ and ${\red 2}$ and the outgoing partons as $\rTH$ and $\rF$.

First consider the channels involving only quarks and anti-quarks.
We start with $q q' \rightarrow q q'$, with $q$ and $q'$ different quark flavors, e.g. 
$u(p_{\rO}) + d(p_{\rT}) \to  u (p_{\rTH}) +  d (p_{\rF}) $.
In this channel, there are two separate color singlet operators for each spin
\begin{align} 
 \cO^{stu}_{\bO \gGG \gGG'} &=
( \bar{\chi}_\rF Y_\rF^\dagger   {\blue \tau^a}\gamma_{\mu}  \gG  Y_\rT \chi_\rT) 
( \bar{\chi}_\rTH Y_\rTH^\dagger {\blue \tau^a}\gamma^{\mu}  \gG' Y_\rO \chi_\rO) \\ 
\cO^{stu}_{\bT \gGG \gGG'} &=
( \bar{\chi}_\rF Y_\rF^\dagger    \gamma_{\mu}  \gG  Y_\rT \chi_\rT) 
( \bar{\chi}_\rTH Y_\rTH^\dagger  \gamma^{\mu}  \gG' Y_\rO \chi_\rO)    \,.
\end{align}
Here, ${\blue \tau^a}$ refer to generators of $\mathrm{SU} (3)$ and the color adjoint index ${\blue a}$ is summed over.
$\gG$ and $\gG'$ label the spin of the operator. 

Since QCD is non-chiral, left- and right-handed fermions
can be thought of as separate species; thus it is simplest to take a basis where $\gG$ and $\gG'$
are either 
${\green P_L} =  \frac{1}{2}\left( 1 -\gamma_5\right)$ or
${\green P_R} = \frac{1}{2}\left( 1 +\gamma_5\right)$, as in~\cite{Chiu:2008vv}.  
Note that the primes on the $\gG$ go with the second contraction, not with the second quark species.
For example, for the $u(p_{\rO}) + d(p_{\rT}) \to  u (p_{\rTH}) +  d (p_{\rF})$ process, 
\begin{equation}
\gGG \gGG' =
\left\{
\begin{array}{c}
{\green LL} = {\green P_L  P_L} \\
{\green LR} = {\green P_L  P_R} \\
\vdots
\end{array}
\right\}
\text{ refers to }
\left\{
\begin{array}{c}
u_L d_L \to u_L d_L \\
u_R d_L \to u_R d_L \\
\vdots
\end{array}
\right\} \,.
\end{equation}
and so on.

For the crossed processes the fermions are contracted differently.  For example, in $qq' \to q'q$,
the fields labels change from $\rF \rT \rTH \rO $ to $\rTH \rT \rF \rO $. The new operators are
\begin{align} 
 \cO^{sut}_{\bO \gGG \gGG'} &=
( \bar{\chi}_\rTH Y_\rTH^\dagger {\blue \tau^a} \gamma_{\mu}\gG  Y_\rT \chi_\rT) 
( \bar{\chi}_\rF Y_\rF^\dagger {\blue \tau^a} \gamma^{\mu}\gG' Y_\rO \chi_\rO) \\ 
\cO^{sut}_{\bT \gGG \gGG'} &=
( \bar{\chi}_\rTH Y_\rTH^\dagger \gamma_{\mu}\gG  Y_\rT \chi_\rT) 
( \bar{\chi}_\rF Y_\rF^\dagger \gamma^{\mu}\gG' Y_\rO \chi_\rO)   \,.
\end{align}
There are $4!$ possible permutations of the labels $\rO \rT \rTH \rF$, 
but the Wilson coefficients for all channels are determined by crossing relations of the Mandelstam invariants
$s$, $t$ and $u$. Therefore, we label the operators with the appropriate permutation of $stu$.
The momentum routing and operators for various crossings are given in Table~\ref{tab:crossqqqq}. 
Our conventions are chosen to agree with~\cite{Ellis:1985er}. 
Note that the  $stu$  channel is defined as $q q' \to q q'$ which has a $t$-channel pole at leading order.

As long as the quarks are distinguishable, only one set of operators is required for each channel. However, there are
additional subtleties when identical particles are present. Consider the process $q q\to q q$. At tree-level,
this process gets a contribution from the $t$-channel and the $u$-channel.
The $t$-channel is like $q q' \to q q'$ and can be matched onto $\cO^{stu}_{\bI \gG}$,
while the $u$-channel is like $q q' \to q' q$, which matches onto $\cO^{sut}_{\bI \gG}$. As long as there is no interference,
both of these operators can be turned on at the same time in the Lagrangian, and the amplitudes added incoherently. In fact,
for the $\gL \gR$ and $\gR \gL$ operators, in which the quark helicities are different, this is the case. 
However, for the $\gL \gL$ and $\gR \gR$, interference is important. 
This can be seen in the effective theory because the operators are not linearly independent. As
was shown in~\cite{Chiu:2008vv},
\begin{equation} \label{Bmix}
 \mathcal{O}^{sut}_{ \bI \gL \gL} = B_{\bI \bJ}  \mathcal{O}^{stu}_{ \bJ \gL \gL}   \,,
\end{equation}
where
\begin{equation}
   B_{\bI \bJ} = 
   \begin{pmatrix}
     -\frac{1}{C_A} & 2 \\ 
     \frac{C_F}{C_A} & \frac{1}{C_A}
   \end{pmatrix}  \,.
\end{equation}
This crossing matrix satisfies $B^2 = 1$, and includes the $-1$ from Fermi statistics. Thus, to treat the $\gL \gL$ 
and $\gR \gR$ channels, only the $\mathcal{O}^{stu}_{ \bI \gL \gL}$ operators are required. We will use this matrix
to extract a single set of Wilson coefficients for the $qq\to qq$ channel in the next section.

For the $gg \to q\bar{q}$ processes, there are three color structures and 16 independent spin/helicity combinations. For the quark 
spins, it makes sense to project out the left- and right-handed states using
the $\gG = {\green P_L}$ or ${\green P_R}$ projectors. For the gluon helicities, we could use a similar projector formalism,
but is is easiest to simply write $\gpm$ as a label for the (incoming) gluon helicities. 
For the $gg \to q\bar{q}$ channel, the operators in QCD are then
\begin{align}
\mathcal{O}^{stu}_{ {\bO} \gpm\, \gpm } &= 
( \bar{q}_{\rTH} {\blue \tau^{a} \tau^{b}} \gG q_{\rF}) (A^{{\blue a}\gpm}_{\rO} A^{{\bb}\,\gpm}_{\rT} ) \nn \\
\mathcal{O}^{stu}_{ \bT \gpm {\gpm}} &= 
( \bar{q}_{\rTH} {\blue \tau^{b} \tau^{a}} \gG q_{\rF}) (A^{{\blue a}\,\gpm}_{\rO} A^{{\bb}\,\gpm}_{\rT} )  \\
\mathcal{O}^{stu}_{ \bTH  \gpm {\gpm}} &= 
( \bar{q}_{\rTH} {\blue \delta^{a b} } \gG q_{\rF})  (A^{{\blue a}\,\gpm}_{\rO} A^{{\bb}\,{\gpm}}_{\rT} )\nn  \,,
\end{align}
which we generically write as
\begin{align}
\mathcal{O}^{stu}_{ {\bI} \gpm\, \gpm } &= 
( \bar{q}_{\rTH} {\blue T^{a b}} \gG q_{\rF}) 
(A^{{\blue a}\gpm}_{\rO} A^{{\bb}\,\gpm}_{\rT} )   \,,
\end{align}
with ${\blue T_\bO}  = {\blue \tau^a \tau^b}$, ${\blue T_\bT}={\blue \tau^b \tau^a}$ and ${\blue T_3} = {\blue \delta^{ab}}$.
The $\gG$ dependence of these operators determines the spin of the out going quarks, and 
since QCD is non-chiral, the amplitude will be independent of $\gG$ and so we leave 
the $\gG$ index on the operators implicit.
The following equation should make the helicity conventions clear.
\begin{equation}
\gpm \gpm =
\left\{
\begin{array}{c}
{\green L} = {\green P_L}; \gp \gp \\
{\green L} = {\green P_L}; \gp \gm \\
{\green R} = {\green P_R}; \gp \gm \\
\vdots
\end{array}
\right\}
\text{ refers to }
\left\{
\begin{array}{c}
g^+ g^+ \to q_L \bar{q}_R \\
g^+ g^- \to q_L \bar{q}_R \\
g^+ g^- \to q_R \bar{q}_L \\
\vdots
\end{array}
\right\} \,.
\end{equation}

It is straightforward to translate these QCD operators into operators in SCET. Gluon fields with associated collinear Wilson lines
are labeled ${\cal A}_{ {\red n} \perp}^\mu$ or equivalently  ${\cal A}_{ {\red n} \perp}^{\gpm}$. 
To make the notation clear, ${\cal A}_{ {\red n} \perp}^{\blue +}$ annihilates an incoming positive 
({\green +}) helicity gluon and creates an outgoing negative ({\green --}) helicity one.
We write ${\cal Y}^{\ba \bb}_{\red n}$ for a soft Wilson line in the adjoint representation in the direction
 ${\red n^\mu}$, as in~\cite{Bauer:2001yt}.
 Then, performing a soft-field redefinition
\begin{equation}
   {\cal A}^{\gpm \ba}_{{\red n} \perp} \to {\cal Y}^{\ba \bb}_{\red n}{\cal A}^{\gpm \bb}_{{\red n}\perp}  
\end{equation}
leads to
\begin{equation} \label{qqggOs}
\mathcal{O}^{stu}_{\bI \gpm\gpm}
 = \left(
 \bar{\chi}_{\rTH}^{\bi} {\cal A}_{\rO \perp}^{\gpm \ba} \gG {\cal A}_{\rT \perp}^{\gpm \bb}\chi^{\bj}_{\rF} 
\right)
\left(
Y^\dagger_{\rTH} {\cal Y}^{\ba \ba'}_\rO {\blue T^{\ba' \bb'}_\bI} {\cal Y}^{\bb \bb'}_\rT Y_{\rF}
\right)^{\bi}_{\ \bj}
\end{equation}
The indices $\ba$ and $\bb$ are adjoint color indices and $\bi$ and $\bj$ are fundamental color indices.
We have also used factorization to pull the soft Wilson lines away from the collinear fields. 
When these operators are squared, the $\ba, \bb, \bi$ and $\bj$ indices in the collinear part (left brackets) 
are contracted with their  counterparts in the adjoint operator through delta functions, 
since the collinear interactions are color diagonal. Therefore, the $\ba, \bb, \bi$ and $\bj$ indices in the soft part 
(right brackets) 
also get contracted with the soft part of the adjoint operator. 
Thus, collinear fields can be completely ignored
when calculating color-related effects. 
This situation is almost identical to the direct photon case, except now with multiple color structures;
more details of how such a factorization arises and how the color and spin indices are contracted can
be found in~\cite{Becher:2009th}.

The crossed operators are constructed in a straightforward way by permuting the indices.
For example, for $qg \to q g$, the operators are
\begin{align}
\mathcal{O}^{tsu}_{ {\bO} \gpm\, \gpm } &=
( \bar{q}_{\rTH} {\blue T^{ab}} \gG q_{\rO})
 ( A_{\rF}^{\ba\, \gmp} A_{\rT}^{\bb\, \gpm} )  \,.
\end{align}
Various crossings and the momentum routing conventions, which again are chosen to agree
with~\cite{Ellis:1985er}, are shown in Table~\ref{tab:crossqqgg}.

For $gg\to gg$, we also label the operators by the gluon helicities, rather than putting in explicit helicity projectors. There
are 8 color channels for this process, 16 spin states, and no crossings.
The matching is more conveniently performed using an over-complete basis of the following 9 operators 
\begin{equation}
\mathcal{O}_{\bI}
^{\gpm \gpm;\gpm \gpm}
  = 
 \Big( 
  {\cal A}_{\rO \perp}^{\gpm \ba}
  {\cal A}_{\rT \perp}^{\gpm \bb}  
  {\cal A}_{\rTH \perp}^{\gpm \bc}
  {\cal A}_{\rF \perp}^{\gpm \bd} 
 \Big)
 \Big(
  {\blue T_\bI^{a' b' c' d'}}
  {\cal Y}^{\ba \ba {\blue '}}_\rO
  {\cal Y}^{\bb \bb {\blue '}}_\rT 
  {\cal Y}^{\bc \bc {\blue '}}_\rTH 
  {\cal Y}^{\bd \bd {\blue '}}_\rF
  \Big)  \,,
\end{equation}
where the color structures are given by
\begin{align}
{\blue T^{a b c d}_{1} } &= \text{Tr}[{\blue \tau^{a} \tau^{b}\tau^{c}\tau^{d} } ] &  
{\blue T^{a b c d}_{6} } &= \text{Tr}[{\blue \tau^{a} \tau^{c}\tau^{b}\tau^{d} } ] \nn \\ 
{\blue T^{a b c d}_{2} } &= \text{Tr}[{\blue \tau^{a} \tau^{b}\tau^{d}\tau^{c} } ] & 
{\blue T^{a b c d}_{7} } &= \text{Tr}[{\blue \tau^{a} \tau^{d} }] \text{Tr}[{\blue \tau^{c}\tau^{b} } ]\nn \\ 
{\blue T^{a b c d}_{3} } &= \text{Tr}[{\blue \tau^{a} \tau^{d}\tau^{c}\tau^{b} } ] &
{\blue T^{a b c d}_{8} } &= \text{Tr}[{\blue \tau^{a} \tau^{b} }] \text{Tr}[{\blue \tau^{c}\tau^{d} } ]\nn \\ 
{\blue T^{a b c d}_{4} } &= \text{Tr}[{\blue \tau^{a} \tau^{d}\tau^{b}\tau^{c} } ] & 
{\blue T^{a b c d}_{9} } &= \text{Tr}[{\blue \tau^{a} \tau^{c} }] \text{Tr}[{\blue \tau^{b}\tau^{b} } ]\nn \\ 
{\blue T^{a b c d}_{5} } &= \text{Tr}[{\blue \tau^{a} \tau^{c}\tau^{d}\tau^{b} } ] 
\label{glue:color}  \,.
\end{align}
While this basis can be reduced to a linearly independent set of 8 operators, there is no need
to do so. As long as the interference effects among the 9 operators are properly accounted for,
there is no difficulty in the effective theory with the basis being overcomplete.

\section{Wilson Coefficients \label{sec:wc}}

The Wilson coefficients $\cC_\bI$ for the operators $\cO_\bI$ are calculated in matching from QCD
to SCET. They are uniquely fixed by requiring that the SCET
Lagrangian reproduce the matrix elements of QCD
order-by-order in $\alpha_s$.  
Since virtual graphs in SCET are scaleless, they vanish in dimensional regularization. Moreover,
since infrared divergences in SCET are the same as in QCD, they cancel in the matching step. Therefore, the matching 
coefficients can be extracted entirely from IR-regulated virtual graphs in QCD. All of the following results
are derived using dimensional regularization and the $\overline{\text{MS}}$ subtraction scheme.

There are a number of results in the literature which are relevant for the calculation of the Wilson coefficients.
The classic work~\cite{Ellis:1985er} presents the cross sections for all the channels separately,
summed over spins and colors. This is not enough information to extract the Wilson coefficients, which we need
at the amplitude level. Refs.~\cite{Chiu:2009ft} and~\cite{Kunszt:1993sd} have
 calculated the necessary diagrams for the $q q \to q q $ channels. 
Refs.~\cite{Fuhrer:2010eu} and~\cite{Kunszt:1993sd} have the $q\bar{q} \to g g$ results. 
For $gg\to gg$ most of the loop amplitudes appear in~\cite{Kunszt:1993sd}, except for a few of the color channels
which can be found in~\cite{Bern:1990cu}. Much of the previous work is written using the spinor helicity formalism. 
In order to simplify the Wilson coefficients, we will convert the spinor helicity products into Mandelstam invariants.

Our momentum convention is $p_\rO+ p_\rT \to p_\rTH+ p_\rF$. We define Mandelstam invariants as
\begin{equation}
  s = (p_\rO + p_\rT)^2,\quad
  t = (p_\rO - p_\rTH)^2,\quad
  u = (p_\rT - p_\rTH)^2  \,.
\end{equation}
This convention agrees with~\cite{Ellis:1985er}, 
but differs from Chiu et al.~\cite{Chiu:2008vv} via $u \leftrightarrow t$.
For all processes, physical kinematics has $s>0$ and $t,u<0$. 
Since the matching
is done at the hard scale we will indiscriminately use
\begin{equation}
  s+t+u=0 
\end{equation}
to simplify the Wilson coefficients.

\subsection{$qq \to qq$ channels}
First, consider the 4-quark processes. There are six channels involving two different flavor quarks (such as $u$ and $d$,
which we call generically $q$ and $q'$)
\begin{equation}
\quad q q' \to q q',\quad q\bar{q}' \to q \bar{q}',\quad q\bar{q} \to \bar{q}' q',
\quad q q' \to q'q, \quad q\bar{q}' \to \bar{q}'q,\quad q\bar{q} \to q' \bar{q}'  \,.
\end{equation}
The first channel we call $stu$. The others are related by crossing symmetries, as shown in 
Table~\ref{tab:crossqqqq}. In addition, there are three identical particle channels
\begin{equation}
q q \to q q, \quad 
q\bar{q} \to q \bar{q},\quad 
q\bar{q} \to \bar{q} q   \,.
\end{equation}
All other channels, such as $\bar{q} \bar{q}' \to \bar{q} \bar{q}'$ and  $\bar{q}\bar{q} \to \bar{q} \bar{q}$,
are the same as one of these by charge conjugation invariance of QCD. Charge conjugation also
swaps $\gL \leftrightarrow \gR$, although this is unimportant since $\gL$ and $\gR$ states never interfere.

\begin{table}[t]
\begin{center}
  \begin{tabular}{|c|c|c||c|c|c|} 
\hline
~~~~$\rO\rT \to \rTH \rF$ & \text{crossing} & \text{operators} &
~~~~$\rO\rT \to \rTH \rF$ & \text{crossing} & \text{operators}\\
\hline
$\phantom{\dfrac{u^2}{t^2_6}}$
$q q' \to q q'$ 
& $s t u$   
& $(\bar{q}_\rF{\blue T}\gG q_\rT) ( \bar{q}_\rTH \gG' {\blue T }q_\rO)$
&$\phantom{\dfrac{u^2}{t^2_6}}$
$q q' \to q' q $ 
& $s u t$   
& $(\bar{q}_\rTH{\blue T}\gG q_\rT) ( \bar{q}_\rF \gG' {\blue T }q_\rO)$
\\
\hline
$\phantom{\dfrac{u^2}{t^2_6}}$
$q \bar{q}' \to q \bar{q}'$    
& $u t s$ 
& $(\bar{q}_\rT{\blue T}\gG q_\rF) ( \bar{q}_\rTH \gG' {\blue T }q_\rO)$
&$\phantom{\dfrac{u^2}{t^2_6}}$
$q \bar{q}' \to \bar{q}' q$    
& $t u s$ 
& $(\bar{q}_\rT{\blue T}\gG q_\rTH) ( \bar{q}_\rF \gG' {\blue T }q_\rO)$
\\
\hline
$\phantom{\dfrac{u^2}{t^2_6}}$
$q \bar{q} \to \bar{q}' q'$    
& $t s u$       
& $(\bar{q}_\rF{\blue T}\gG q_\rTH) ( \bar{q}_\rT \gG' {\blue T }q_\rO)$
&  $\phantom{\dfrac{u^2}{t^2_6}}$ 
$q \bar{q} \to q' \bar{q}'$    
& $u s t$       
& $(\bar{q}_\rTH{\blue T}\gG q_\rF) ( \bar{q}_\rT \gG' {\blue T }q_\rO)$
\\
\hline
  \end{tabular}
\end{center}
\caption{Crossing relations for the 4-quark channels.  
For example, the Wilson coefficients for the $tsu$
    channel are obtained from the $stu$ channel via 
   $C^{\gG}_{\bI; tsu}(s,t,u) = C^{\gG}_{\bI; stu}(t,s,u)$.
\label{tab:crossqqqq}
}
\end{table}

For all the non-identical particle channels, the Wilson coefficients can be written as
 in the general form
\begin{align} \label{WCform}
\cC_{{\bO}}^{\gL \gL}(s,t,u) &= 2g^2\frac{s}{t}
\left\{ 
1 + \frac{\alpha_s}{4 \pi} 
\left[-2C_FL(t)^2 + X_{\bO}(s,t,u)L(t) +Y+\phantom{\frac{1}{2}}\!\!\!\! (\frac{1}{2}C_A-2C_F) Z(s,t,u)\right]
\right\}  \, , \nn \\
\cC_{{\bO}}^{\gL \gR}(s,t,u) &=2g^2\frac{u}{t}
\left\{ 
1 + \frac{\alpha_s}{4 \pi} 
\left[-2C_F L(t)^2 + X_{\bO}(s,t,u)L(t) +Y+\phantom{\frac{1}{2}}\!\!\!\! (2C_F-C_A) Z(u,t,s)\right]
\right\}  \, , \nn \\
\cC_{{\bT}}^{\gL \gL}(s,t,u) &= 2g^2\frac{s}{t}
 \left\{  
\frac{\alpha_s}{4 \pi} 
\left[ X_{\bT}(s,t,u)L(t)-\frac{C_F}{2C_A} Z(s,t,u)\right]
\right \} \, , \nn \\
\cC_{{\bT}}^{\gL \gR}(s,t,u) &=  2g^2\frac{u}{t}
 \left\{ \frac{\alpha_s}{4 \pi} 
\left[ X_{\bT}(s,t,u)L(t) +\frac{C_F}{2C_A} Z(u,t,s)\right]
\right\} \nn \, ,
\end{align}
with
\begin{align}
X_\bO(s,t,u) &= 
\phantom{\frac{C_F}{C_A}\!\!\!\!\!\!\!\!} 6 C_F - \beta_0 + 8 C_F [L(s) - L(u)] - 2C_A[2L(s)-L(t)-L(u)]
\\
X_\bT(s,t,u) &= 
\frac{2C_F}{C_A} [L(s)-L(u)]
\\
Y &=C_A\left(\frac{10}{3} + \pi^2\right) + C_F\left(\frac{\pi^2}{3} - 16\right) + \frac{5}{3}\beta_0\\
Z(s,t,u) &= \frac{t}{s}\left(\frac{t+2 u}{s}[L(u) - L(t)]^2 +2[ L(u)-L(t)] + \pi^2\frac{t + 2 u }{s}\right)   \,,
\end{align}
where 
\begin{equation} \label{b0def}
\beta_0 = \frac{11}{3} C_A - \frac{2}{3} n_f  \,.
\end{equation}
and $C_A=N=3$ and $C_F=\frac{N^2-1}{2N} =\frac{4}{3}$, with $N=3$ the number of colors.
The function $L(x)$ is defined by $L(x) = \log(|x|/\mu^2) - i \pi \theta(x)$. For massless $2\to 2$ scattering, $s>0$
and $t,u<0$, so
\begin{align}
  L(t) &= \log\frac{-t}{\mu^2},\nn \\
  L(u) &= \log\frac{-u}{\mu^2}, \label{Ldef}\\
  L(s) &= \log\frac{s}{\mu^2} - i \pi \nn  \,.
\end{align}
It is important to cross the Wilson coefficients {\it before} extracting the imaginary parts from these logarithms.

The hard function $H_{\bI\bJ} =\sum_\gG  \cC^{\gG }_{\bI} \cC^{\gG \star}_{\bJ }$ is then, up to order $g^4 \alpha_s$,
\begin{align}
  H_{\bO \bO}(s,t,u) &= 8g^4\frac{s^2+u^2}{t^2} + 8g^4\frac{\alpha_s}{4\pi} 
\Big[\frac{s^2+u^2}{t^2} \left( -4 C_F L(t)^2 + 2 X_\bO (s,t,u) L(t)+2 Y\phantom{\frac{a}{b}}\!\!\right)\\\nn
&
+\frac{s^2}{t^2}(C_A-4C_F) Z(s,t,u) -\frac{u^2}{t^2} (2C_A-4C_F) Z(u,t,s)\Big]\\\nn
H_{\bO \bT}(s,t,u) &= 8g^4\frac{\alpha_s}{4\pi}
 \Big[\frac{s^2+u^2}{t^2} X_\bT (s,t,u) L(t)
-\frac{s^2}{t^2}\frac{C_F}{2C_A}Z(s,t,u) +\frac{u^2}{t^2}\frac{C_F}{2C_A} Z(u,t,s)\Big]\\ \nn
H_{\bT \bO}(s,t,u) &= H_{\bO \bT}(s,t,u) \\
H_{\bT \bT}(s,t,u) &= 0\, . \nn
\end{align}

 As a check, we can compare $\la{\mathcal M}_{(2\to2)}\ra^2 =\frac{1}{N_{\text{init}}} \text{Tr}[ H S^\mathrm{tree} ]$
to the $2\to 2$ matrix-elements-squared in QCD, as in 
Eq.~\eqref{qcdcheck}. 
For example, in  the $q\bar{q}\to \bar{q}' q'$ channel ($tsu$), 
at tree-level
\begin{align}
  \cC_{\bO}^{\gL \gL} = \cC_{\bO}^{\gR \gR} = 2g_s^2 \frac{t}{s},\quad
  \cC_{\bO}^{\gL \gR} = \cC_{\bO}^{\gR \gL} = 2g_s^2 \frac{u}{s},\quad
\cC_\bT^{\gG}=0
~~~~~\qquad (q\bar{q}\to \bar{q}' q')  \,.
\end{align}
So,
 \begin{equation}
   H^{\mathrm{tree}}_{\bI \bJ} =8 g_s^4\begin{pmatrix}
1\, & 0 \\ 
 0\, & 0\\ 
   \end{pmatrix} 
\left(\frac{t^2 + u^2}{s^2}\right)
~~~~~\qquad (q\bar{q}\to \bar{q}' q')  \,.
 \end{equation}
At tree-level, the soft function is
\begin{equation}
   S^{\text{tree}}_{\bI \bJ} =
 \begin{pmatrix}
\frac{1}{2} C_A C_F  & 0 \\ 
 0 & C_A^2\\ 
   \end{pmatrix}  \,.
\end{equation}
For this channel, $N_{\text{init}} = 4N^2$ giving a final result
\begin{equation}
\la{\mathcal M}_{(q\bar{q}\to \bar{q}'q')}\ra^2=\frac{1}{4 N^2}  g_s^4
 4 C_A C_F\frac{t^2 + u^2}{s^2}+ \cdots  \,.
\end{equation}
The result is the correct tree-level matrix-element-squared for parton level $q \bar{q} \to q' \bar{q}'$ 
scattering in QCD. Using the NLO hard function, the result agrees with~\cite{Ellis:1985er}.

For identical particles, the Wilson coefficients should be included only for linearly independent operators. 
This is not strictly necessary, but avoids having to compute interference effects.
As discussed
in Section~\ref{sec:operators}, for $q q \to q q$ in the $\gL \gL$ spin configuration, the operators 
$\cO^{stu}_{\bI \gL \gL}$ and $\cO^{sut}_{\bI \gL \gL}$ are not independent. To correct for this, we can match onto the
operators separately, and then use Eq.~\eqref{Bmix} to write
\begin{align}
 \cC_{\bI}^{\gL \gL\gL\gL} &= \cC_{\bI}^{\gR \gR\gR\gR}=  \cC_{\bI}^{\gL \gL}(s,t,u)+ B_{\bI \bJ}\cC_{\bJ}^{\gL \gL}(s,u,t)
\qquad (qq\to qq)  \,.
\end{align} 
We have added more spin labels to make the counting of independent states more transparent.
We can also cross the $\gR \gL$ channel into the $stu$ basis so that it can be evolved with the $uts$ RG kernel (see Section~\ref{sec:RGEqq}
below). 
So we define for the $qq \to qq$ channel
\begin{align}
 \cC_{\bI}^{\gL \gR \gL \gR} &=  \cC_{\bI}^{\gR \gL \gR \gL} = \cC_{ \bI}^{ \gL \gR}(s,t,u)
\:\:\:\quad \qquad (qq\to qq) \, ,\\
 \cC_{\bI}^{\gL \gR \gR \gL} &= \cC_{\bI}^{\gR \gL \gL \gR} =   B_{\bI \bJ}  \cC_{ \bJ}^{\gL \gR} (s,u,t)
\qquad (qq\to qq)  \,.
\end{align}
These contributions can then be combined incoherently. Summing over the six contributions, the tree-level
hard function is then
\begin{align}
  H^{\text{tree}}_{\bI \bJ} &= \frac{8 g^4}{C_A^2 t^2 u^2}
\begin{pmatrix}
t^4 + C_A^2 u^4 + s^2(t-C_A u)^2\quad & - C_F (t^4 + s^2 (t^2-C_A t u) ) \\
-C_F (t^4 +s^2 (t^2-C_A tu)) & C_F^2 t^2 (s^2+ t^2)
\end{pmatrix}  \,.
\qquad (qq\to qq)\nn
\end{align}
Multiplying by the tree-level soft function and dividing by $4 N^2$ to average
over initial states gives the correct tree-level $q q \to q q$ matrix element in QCD
\begin{equation}
  |\cM_{(qq\to qq)}|^2 = \frac{4}{9}\left(\frac{s^2+u^2}{t^2} + \frac{t^2 + u^2}{s^2}\right) - \frac{8}{27}\frac{u^2}{s t}+\cdots
  \,.
\end{equation}
Using the NLO hard function, $|\cM|^2$ also agrees with the NLO result in full QCD~\cite{Ellis:1985er}.

For $q\bar{q} \to q\bar{q}$, with the momentum convention in Table~\ref{tab:crossqqqq},
we need the $uts$ and $ust$ operators. As
for $q q \to q q$ the $\gL\gR$ spins add incoherently and should be treated as separate channels,
while the $\gL\gL$ operators interfere. In this case, the same matrix applies. Thus, we need
\begin{align}
 \cC_{\bI}^{\gL \gL \gL \gL} &=  \cC_{\bI}^{\gR \gR \gR \gR}= \cC_{ \bI}^{\gL \gL}(u,t,s)+ B_{\bI \bJ}  \cC_{ \bJ }^{\gL \gL}(u,s,t)
\qquad (q\bar{q}\to q\bar{q})\\
 \cC_{\bI}^{\gL \gR \gL \gR} &=\cC_{\bI}^{\gR \gL \gR \gL} =  \cC_{ \bI}^{ \gL \gR} (u,t,s) 
~~~~~~~~~~~~~~~~~~
\:\:\:\quad\qquad (q\bar{q}\to q\bar{q})\\ 
 \cC_{\bI}^{\gR \gL \gL \gR} &=\cC_{\bI}^{\gL \gR \gR \gL} = B_{\bI \bJ} \cC_{\bJ}^{ \gL \gR} (u,s,t)  \,.
~~~~~~~~~~~~~~~~~ \qquad (q\bar{q}\to q\bar{q})
\end{align}
These four contributions can then be combined incoherently into a single hard function, as for $qq\to qq$. For 
$q\bar{q} \to \bar{q} q$, the channels are $tsu$ and $tus$.

\subsection{$gg \to q \bar{q}$ channels}
Next, we give the Wilson coefficients for the $gg \to q\bar{q}$ channel and its crossings. 
There are again six crossings
\begin{equation}
gg \to q\bar{q}, \quad
qg \to qg, \quad 
\bar{q}g \to \bar{q}g, \quad 
gg \to \bar{q} q, \quad
qg \to gq, \quad 
\bar{q} g \to g \bar{q}  \,.
\end{equation}
The relevant operators, momentum conventions, and crossing relations are given in Table~\ref{tab:crossqqgg}.
The Wilson coefficients for $q\bar{q} \to gg$ are the same as $gg \to q\bar{q}$, but one must keep in mind
the different number of color and spins in the initial state for a spin-averaged cross section.
The amplitudes for quarks are the same as the amplitudes for anti-quarks, but with 
opposite gluon helicities,
$\gp \leftrightarrow \gm$ and opposite spins $\gL \leftrightarrow \gR$.
Since we will be mostly interested in the hard function $H_{\bI \bJ}$ which sums over helicities, the convention we choose
for $\gpm$ and $\gL/\gR$ is unimportant.

\begin{table}[t] 
\begin{center}
  \begin{tabular}{|c|c|c||c|c|c|} 
\hline
$\rO\rT \to \rTH \rF$ & \text{crossing} & \text{operators} &
$\rO\rT \to \rTH \rF$ & \text{crossing} & \text{operators}\\
\hline
$\phantom{\dfrac{u^2}{t^2_6}}$
$g g \to q \bar{q}$
& $s t u$   
& $(\bar{q}_{\rTH} {\blue T^{a b}} \gG q_{\rF})(A^{{\blue a}\gpm}_{\rO} A^{{\bb}\,\gpm}_{\rT} )$
& 
$ g g \to \bar{q} q  $
& $s u t$   
& $(\bar{q}_{\rF} {\blue T^{a b}} \gG q_{\rTH})(A^{{\blue a}\gpm}_{\rO} A^{{\bb}\,\gpm}_{\rT} )$
\\
\hline
$\phantom{\dfrac{u^2}{t^2_6}}$
$\bar{q} g \to g \bar{q}$ 
& $u t s$   
& $(\bar{q}_{\rO} {\blue T^{a b}} \gG q_{\rF})(A^{{\ba} \gmp}_{\rTH} A^{{\bb}\,\gpm}_{\rT} )$
&
$\bar{q} g \to \bar{q} g$ 
& $t u s$   
& $(\bar{q}_{\rO} {\blue T^{a b}} \gG q_{\rTH})(A^{{\ba} \gmp}_{\rF} A^{{\bb}\,\gpm}_{\rT} )$
\\
\hline
$\phantom{\dfrac{u^2}{t^2_6}}$
$q g \to q g$
& $t s u$   
& $(\bar{q}_{\rTH} {\blue T^{a b}} \gG q_{\rO})(A^{{\ba} \gmp}_{\rF} A^{{\bb}\,\gpm}_{\rT} )$
& 
$q g \to g q$ 
& $u s t$   
& $(\bar{q}_{\rF} {\blue T^{a b}} \gG q_{\rO})(A^{{\ba} \gmp}_{\rTH} A^{{\bb}\,\gpm}_{\rT} )$
\\
\hline
  \end{tabular}
\caption
  {
    Crossing relations for the $gg \to q\bar{q}$ channels.  
\label{tab:crossqqgg}
  }
\end{center}
\end{table}

The Wilson coefficients $\cC_{\bI}^{\gpm \gpm; \gpm\gpm}=\cC_{\bI}^{\gl_\rO \gl_\rT; \gl_\rTH \gl_\rF}$
have 4 helicity labels ${\green \lambda}_{\red n}$ and one color label, $\bI$.  Also, helicity is conserved
for massless quarks along a fermion line,  so the Wilson coefficients vanish unless
 ${\green \lambda}_{\red 4} = - {\green \lambda}_{\red 3}$. Parity invariance implies that
\begin{equation}
\label{parity}
\cC_{\bI}^{\gl_\rO \gl_\rT; \gl_\rTH \gl_\rF}
=
\cC_{\bI}^{-\gl_\rO -\gl_\rT;- \gl_\rTH -\gl_\rF}  \,.
\end{equation}
Therefore, we only give expressions for half of the Wilson coefficients $\cC_{\bI}^{\gl_\rO \gl_\rT} \equiv \cC_{\bI}^{\gl_\rO \gl_\rT; \gp \gm}$
for $gg \to q\bar{q}$ with fixed quark helicities $\gl_\rTH ={\green +}$ and $\gl_\rF ={\green -}$.  The other half
is obtained through Eq.~\eqref{parity}.

The Wilson coefficients for the 4 independent spins and 3 independent colors are
\begin{align}
\cC_{{\bO}}^{\gm \gp}(s,t,u) &=  2g^2\frac{\sqrt{t u}}{s}
\left\{ 1 + \frac{\alpha_s}{4 \pi} \left[
-(C_A+C_F)L(s)^2 + V_\bO(s,t,u)L(s) + W_\bO (s,t,u)\right]
\right\}  \\
\cC_{{\bO}}^{\gp \gm}(s,t,u) &=  2g^2\frac{u}{s}\sqrt{\frac{u}{t}}
\left\{ 1 + \frac{\alpha_s}{4 \pi} \left[
-(C_A+C_F)L(s)^2 + V_\bO(s,t,u)L(s) + W_\bT (s,t,u)\right]
\right\} \nn \\
\cC_{{\bO}}^{\gp \gp}(s,t,u) &= \cC_{{\bO}}^{\gm \gm}(s,t,u) = 2g^2\sqrt{\frac{u}{t}}
\frac{\alpha_s}{4 \pi} W_\bTH(s,t,u)\nn\\
\cC_{{\bT}}^{\gm \gp}(s,t,u) &= 
\cC_{{\bO}}^{\gp \gm}(s,u,t)\nn  \\
\cC_{{\bT}}^{\gp \gm}(s,t,u) &= 
\cC_{{\bO}}^{\gm \gp}(s,u,t) \nn \\
\cC_{{\bT}}^{\gp \gp}(s,t,u) &= \cC_{{\bT}}^{\gm \gm}(s,t,u) =
\cC_{{\bO}}^{\gp \gp}(s,u,t)\nn  \\
\cC_{{\bTH}}^{\gm \gp}(s,t,u) &= 2g^2\sqrt{\frac{t}{u}}  \frac{\alpha_s}{4 \pi} 
\left\{ V_{\bT}(s,t,u) L(s) +W_{\bF}(s,t,u)\right\} \nn\\
\cC_{{\bTH}}^{\gp \gm}(s,t,u) &=\cC_{{\bTH}}^{\gm \gp}(s,u,t) \nn\\
\cC_{{\bTH}}^{\gp \gp}(s,t,u) &=\cC_{{\bTH}}^{\gm \gm}(s,t,u)=0 \nn  \,,
\end{align}
where
\begin{align}
  W_\bO (s,t,u) &= (C_A-C_F)\frac{s}{u} \Big( [L(s) - L(t)]^2 + \pi^2\Big) + C_A - 8 C_F + \left( \frac{7C_A+C_F}{6}\right)\pi^2\\
W_\bT(s,t,u) &= \left(-C_F \frac{s^3}{u^3} - C_A\frac{t^3+u^3-s^3}{2u^3} \right)
\Big( [ L(s) - L(t)]^2 + \pi^2 \Big) \nn \\
&~~~
+\left(2C_A \frac{t s}{u^2} + C_F \frac{s(2s-u)}{u^2}\right)[ L(t) - L(s)] 
+C_F \frac{t-7u}{u} - C_A\frac{t}{u} + \left(\frac{7 C_A + C_F}{6}\right)\pi^2 \nn\\
W_\bTH (s,t,u) &= 2C_F-2C_A - \frac{2t}{3s} (C_A- n_f) \nn \\
W_\bF(s,t,u) &= -\frac{3u}{4t} [ L(s)-L(u)]^2 - [L(s) - L(t)][L(s)-L(u)] + \frac{3\pi^2}{2} \frac{u^2}{t s}\nn\\
V_\bO(s,t,u) &= 3 C_F - 2 C_A[ L(t)-L(s)]\nn\\
V_\bT(s,t,u) &= [L(s)-L(u)]+\frac{t}{s}[L(t)-L(u)]\nn  \,.
\end{align}

With these Wilson coefficients, 
it is easy to compute that the hard function for $gg \to q\bar{q}$ is
\begin{equation}
  H(s,t,u) = 8 g^4 \begin{pmatrix}
\frac{u^3+u t^2}{s^2 t} & \frac{t^2 +u^2}{s^2} & 0 \\
\frac{t^2 + u^2}{s^2} & \frac{t(t^2+u^2)}{s^2 u} & 0 \\
0 & 0 & 0 
  \end{pmatrix} +  8 g^4 \frac{\alpha_s}{4\pi} \Re\left[H_{\bI \bJ}^{(NLO)}\right] + \cdots  \,,
\end{equation}
where $\Re[x]$ denotes the real part of $x$,
with
\begin{align}
H_{\bO \bO}^{(NLO)} &= \frac{t^2u+u^3}{s^2t}\Big[(-C_A-C_F) L(s)^2 + V_\bO(s,t,u)L(s) \Big] 
+ \frac{t u}{s^2} W_\bO (s,t,u) + \frac{u^3}{s^2 t} W_\bT(s,t,u)\nn \\
H_{\bT \bT}^{(NLO)} &=  \frac{t^3+tu^2}{s^2u}\Big[(-C_A-C_F) L(s)^2 + V_\bO(s,u,t)L(s) \Big] 
+ \frac{t u}{s^2} W_\bO (s,u,t) + \frac{t^3}{s^2 u} W_\bT(s,u,t)\nn \\
H_{\bO \bT}^{(NLO)} &=  \frac{t^2+u^2}{2s^2}\Big[(-2C_A-2C_F) L(s)^2 +V_\bO(s,t,u) L(s) + V_\bO(s,u,t) L(s)\Big] \\
&~~~~~~~~ + \frac{u^2}{2s^2}(W_\bO(s,u,t)+ W_\bT(s,t,u)) + \frac{t^2}{2s^2} (W_\bO(s,t,u) + W_\bT(s,u,t)) \nn \\
H_{\bO \bTH}^{(NLO)} &=  \frac{t}{2s} V_\bT(s,t,u) L(s) + \frac{u^2}{2st} V_\bT(s,u,t)L(s) + \frac{t}{2s} W_\bF(s,t,u) + \frac{u^2}{2st} W_\bF(s,u,t) \nn\\
H_{\bT \bTH}^{(NLO)} &=  \frac{t^2}{2su} V_\bT(s,t,u) L(s) + \frac{u}{2s} V_\bT(s,u,t)L(s) + \frac{t^2}{2su} W_\bF(s,t,u) + \frac{u}{2s} W_\bF(s,u,t) \nn\\
H_{\bTH \bTH}^{(NLO)} &=0\nn \, ,
\end{align}
and $H_{\bJ \bI} = H_{\bI \bJ}$.

The tree-level soft function in this color basis is
\begin{equation}
   S^{\text{tree}}_{\bI \bJ} = 
   \begin{pmatrix}
        C_A C_F^2 ~~ &  -\frac{1}{2}C_F       ~~     & C_A C_F  ~~        \\
        -\frac{1}{2} C_F   &  C_A C_F^2  & C_A C_F          \\ 
        C_A C_F           &  C_A C_F                & 2 C_F C_A^2  
   \end{pmatrix}  \,.
\end{equation}
Using these results, we can check the $2\to 2$ cross section. 
For $gg \to q\bar{q}$,
\begin{align}
  \la \cM_{(gg \to q\bar{q})}\ra^2 &= \frac{1}{4\times8^2} H_{\bI \bJ}(s,t,u) S^\text{tree}_{\bJ \bI}\\
 &=   g^4\frac{C_F}{32} \frac{t^2+u^2}{s^2} \left( C_A C_F \frac{t^2+u^2}{t u} -1\right) + \cdots  \,,
\end{align}
which is the correct tree level result. The 1-loop hard function when combined with
tree-level soft function reproduces the NLO cross section in~\cite{Ellis:1985er}.


\subsection{$gg\to gg$ channel}
Finally, we give the Wilson coefficients for $gg\to gg$. As discussed in Section~\ref{sec:operators}, there are 8 independent color
structures, but we match to an overcomplete basis of 9 color structures. There are 16 possible helicity amplitudes for each color
structure, giving 144 matching coefficients. Fortunately, there is a great deal of symmetry for this process which relates the
various helicity and color subamplitudes. Parity allows us to give the results for half of the helicities,
with the other half obtained via Eq.~\eqref{parity}. In this section, we will number the helicities $\gG = 1\cdots 16$,
with the correspondence given in Table~\ref{tab:Helicity}.
\begin{table}
\begin{center}
\begin{tabular}{|c|c||c|c|}
\hline
$ \gG  $ & $(\gl_\rO \gl_\rT \to \gl_\rTH \gl_\rF)$  & $ \gG  $ & $(\gl_\rO \gl_\rT \to \gl_\rTH \gl_\rF)$   \\ \hline
$ 1    $ & $( + + , + + )$  & $ 9  $ & $(- + , + +)$   \\
$ 2    $ & $( - - , - - )$  & $ 10 $ & $(+ - , - -)$   \\
$ 3    $ & $( - + , - + )$  & $ 11 $ & $(+ - , + +)$   \\
$ 4    $ & $( + - , + - )$  & $ 12 $ & $(- + , - -)$   \\
$ 5    $ & $( + - , - + )$  & $ 13 $ & $(+ + , - +)$   \\
$ 6    $ & $( - + , + - )$  & $ 14 $ & $(- - , + -)$   \\
$ 7    $ & $( - - , + + )$  & $ 15 $ & $(+ + , + -)$   \\
$ 8    $ & $( - + , + + )$  & $ 16 $ & $(- - , - +)$   \\
\hline
\end{tabular}
\end{center}
\caption{The 16 helicity structures ($\gG$) for the $gg\to gg$ channel.\label{tab:Helicity}}
\end{table}

Only the first 6 color structures in Eq.~\eqref{glue:color} are non-zero at tree-level, 
and then only for 6 of the possible 16 helicity amplitudes. The tree-level amplitudes, 
$\cM_{\bI}^{\gG}$ are given in Table~\ref{TGlue1}.
At one loop, all 16 helicity amplitudes are non-zero for all color channels.
The NLO matching coefficients are given by the following formula 
\begin{equation}
\label{Cglue1}
\cC_{\bI}^{\gG} 
= \left\{
  \begin{array}{lll} 
    \ds
    4g^2 \mathcal{M}_{\bI}^{\gG}
    \left( 1 + \frac{\alpha_s}{4\pi} \mathcal{Q}_{\bI}^{\gG} \right)  &,\quad  {\blue       I = 1 \cdots 6} 
                                                                          &,\quad  {\green \Gamma = 1 \cdots 6}
    \\[10pt]
    \ds
    4g^2 \frac{\alpha_s}{4\pi} \mathcal{Q}_{\bI}^{\gG}                 &,\quad  {\blue I = 7, 8 , 9}
                                                                          &,\quad  {\green \Gamma = 1 \cdots 6}
    \\[10pt]
    \ds
    4g^2 \frac{\alpha_s}{4\pi} \mathcal{Q}_{\bI}^{\gG}                 &,\quad  {\blue I = 1 \cdots 9} 
                                                                          &,\quad  {\green \Gamma = 7 \cdots 16} 
  \end{array}
  \right.  \,,
\end{equation} 
where the expressions for $\mathcal{Q}_{\bI}^{\gG}$ are given in Table~\ref{TGlue2a}.
The quantities $\mathcal{Q}_{\bI}^{\gG}$ are written in terms of 
$\mathcal{A}$, $\mathcal{B}$, and $\mathcal{F}$, which are as follows:
\begin{align}
\mathcal{A}(s,t,u)
&= 
   -2 C_A L(u)^2 
   +\Big( - 2C_A [L(s)-L(u)] + \beta_0 \Big) L(u) 
   +\left( \frac{4\pi^2}{3} - \frac{67}{9} \right) C_A   
   +\frac{10}{9}n_f 
\nn\\
\mathcal{B}(s,t,u)
&=
\mathcal{A}(s,t,u)    + \beta_0 \frac{u}{t}[L(u) - L(s)]
    - \frac{3n_f}{2} \frac{su}{t^2} \Big( [L(u) - L(s)]^2 + \pi^2 \Big)
 \\ 
&    + (C_A-n_f)    \frac{su}{t^2} 
    \left[ \frac{s-u}{t} [L(u) - L(s)] + \left(\frac{su}{t^2} - 2 \right)
   \Big( [L(u) - L(s)]^2 + \pi^2 \Big) - 1 \right]
\nn\\
\mathcal{F}(s,t,u)
&= 
\frac{1}{C_A} \left(
\frac{s^2}{t u}\mathcal{B}(t,s,u)
+
\frac{s^2}{t u}\mathcal{B}(u,s,t)
+
\frac{2s}{u} \mathcal{A}(s,t,u)
+
\frac{2s}{t}  \mathcal{A}(s,u,t)
\right)\nn  \,.
\end{align}  
\begin{table}
\begin{center}
  \begin{tabular}{|c|c|c|c||c|c|c|c|}
\hline
${\mathcal{M}_{\bI}^{\gG}}$ &   $\gG = {\green 1,2}      $ & $\quad{\green 3,4}\quad$ & $\quad{\green 5,6}\quad$ & 
                                &   $\quad {\green 1,2} \quad$ & $\quad{\green 3,4}\quad$ & $\quad{\green 5,6}\quad$ 
\\
\hline
${I = \blue 1}$ & $\dfrac{s}{u}$    & $\dfrac{u}{s}    $ & $\dfrac{t^2}{su} $ & 
${\blue 4}$     & $\dfrac{s^2}{tu}$ & $\dfrac{u}{t}    $ & $\dfrac{t}{u}    $ \\[10pt] 
${\blue 2}$     & $\dfrac{s}{t}$    & $\dfrac{u^2}{st} $ & $\dfrac{t}{s}    $ &
${\blue 5}$     & $\dfrac{s}{t}$    & $\dfrac{u^2}{st} $ & $\dfrac{t}{s}    $ \\[10pt] 
${\blue 3}$     & $\dfrac{s}{u}$    & $\dfrac{u}{s}    $ & $\dfrac{t^2}{su} $ &
${\blue 6}$     & $\dfrac{s^2}{tu}$ & $\dfrac{u}{t}    $ & $\dfrac{t}{u}    $ \\[10pt] 
\hline
  \end{tabular}
\end{center}
\caption{Tree-level matching coefficients, $\mathcal{M}_{{\bI} }^{\gG}$, for the $gg\to gg$ channel. 
\label{TGlue1} }
\end{table}
\begin{table}
\begin{center}
  \begin{tabular}{|c|c c c|c c|}
\hline
  $\mathcal{Q}_{\bI}^{\gG}$ 
& ${\green \Gamma = 1,2}  $ 
& ${\green 3,4}           $
& ${\green 5,6}           $
& ${\green 7}             $
& ${\green 8-16}             $
\\
\hline
${I = \blue 1}$ & $\mathcal{A}( s, t, u ) $    
                & $\mathcal{A}( u, t, s ) $ 
                & $\mathcal{B}( s, t, u ) $ 
                & $\frac{1}{3} (C_A - n_f) $ 
                & $-\dfrac{t^2}{3su}(C_A - n_f) $ 
                                            \\[10pt]
${\blue 2}$     & $\mathcal{A}( s, u, t ) $    
                & $\mathcal{B}( t, u, s ) $ 
                & $\mathcal{A}( t, u, s ) $
                & $\frac{1}{3} (C_A - n_f)$ 
                & $-\dfrac{u^2}{3st} (C_A - n_f)$ 
                                            \\[10pt]
${\blue 3}$     & $\mathcal{A}( s, t, u ) $    
                & $\mathcal{A}( u, t, s ) $ 
                & $\mathcal{B}( u, t, s ) $ 
                & $\frac{1}{3} (C_A - n_f)$ 
                & $-\dfrac{t^2}{3su} (C_A - n_f)$ 
                                            \\[10pt]
${\blue 4}$     & $\mathcal{B}( u, s, t ) $    
                & $\mathcal{A}( u, s, t ) $ 
                & $\mathcal{A}( t, s, u ) $
                & $\frac{1}{3} (C_A - n_f)$ 
                & $-\dfrac{s^2}{3tu} (C_A - n_f)$ 
                                            \\[10pt]
${\blue 5}$     & $\mathcal{A}( s, u, t ) $    
                & $\mathcal{B}( s, u, t ) $ 
                & $\mathcal{A}( t, u, s ) $
                & $\frac{1}{3} (C_A - n_f)$ 
                & $-\dfrac{u^2}{3st} (C_A - n_f)$ 
                                            \\[10pt]
${\blue 6}$     & $\mathcal{B}( t, s, u ) $    
                & $\mathcal{A}( u, s, t ) $ 
                & $\mathcal{A}( t, s, u ) $
                & $\frac{1}{3} (C_A - n_f)$ 
                & $-\dfrac{s^2}{3tu} (C_A - n_f)$ 
                                            \\[10pt]
\hline
${\blue I = 7,8,9}$ & $\mathcal{F}( s, t, u ) $    
                & $\mathcal{F}( u, s, t ) $ 
                & $\mathcal{F}( t, s, u ) $
                & $2                      $ 
                & $-2                      $ 
                                            \\[2pt]
\hline
  \end{tabular}
\end{center}
\caption{
The expressions for $\mathcal{Q}_{\bI}^{\gG}$ for the indicated helicity structures. 
 \label{TGlue2a} }
\end{table}
With these Wilson coefficients, we compute the hard function for $gg \to gg$ 
is
\begin{equation}
H_{\bI \bJ} = 32g^4 \frac{(s^4 + t^4 + u^4 ) }{ s^2 t^2 u^2} 
\left(
\begin{array}{cccccc|c}
 t^2 & t u & t^2          & s t & t u & s t &  \\
 t u & u^2 & t u          & s u & u^2 & s u &  \\
 t^2 & t u & t^2          & s t & t u & s t &  \\
 s t & s u & s t          & s^2 & s u & s^2 & 0_{6\times 3} \\
 t u & u^2 & t u          & s u & u^2 & s u &  \\
 s t & s u & s t          & s^2 & s u & s^2 &  \\ \hline
     &     & 0_{3\times 6} &     &     &     & 0_{3\times 3}
\end{array}
\right) + g^4 \frac{\alpha_s}{4\pi}\Re[  H_{\bI \bJ}^{( NLO )} ]  \,.
\end{equation}
The 1-loop hard function, $H_{\bI \bJ}^{(NLO)}$ is complicated and is therefore not written 
explicitly.

The tree-level soft function in this color basis is
\begin{equation}
S^{\text{tree}}_{\bI \bJ} =
\frac{C_F}{8C_A}
\left(
\begin{array}{ccccccccc}
 a & b & c & b & b & b & d  & d & -e \\
 b & a & b & b & c & b & -e & d &  d \\
 c & b & a & b & b & b & d  & d & -e \\
 b & b & b & a & b & c & d  & -e & d \\
 b & c & b & b & a & b & -e &  d & d \\
 b & b & b & c & b & a & d  & -e & d \\
 d  & -e &  d &  d & -e &  d &   d e & e^2 & e^2 \\
 d  & d  &  d & -e &  d & -e & e^2 & d e  & e^2 \\
 -e & d  & -e &  d &  d &  d & e^2 & e^2 & d e
\end{array}
\right)  \,,
\end{equation}
with
\begin{align}
a &= 3+ C_A^2  & b &= 3 - C_A^2    & c &= C_A^4 - 3C_A^2 + 3 \nonumber \\
d &= 2C_A^2C_F & e &= C_A   \,.
\end{align}
Combining $H_{\bI \bJ}$ with $S^{\text{tree}}_{\bJ \bI}$ and using $N = 3$ gives
\begin{equation}
  | {\mathcal{M}}_{(gg \to gg)}|^2 = \frac{1}{4\times 8^2}H_{\bI \bJ} S^\text{tree}_{\bJ \bI} =  \frac{ C_A^3 C_F g^4}{32} \frac{(s^2+t^2+u^2)(s^4+t^4+u^4)}{s^2u^2 t^2}  + \cdots  \,,
\end{equation}
which is the correct tree-level matrix-element-squared for $gg \to gg$.  
As with the other channels, the 1-loop hard function for $gg\to gg$, when combined with the tree-level soft function, reproduces the NLO 
cross section in full QCD~\cite{Ellis:1985er}.

\section{Renormalization Group Evolution \label{sec:RGE}}
The renormalization group evolution of Wilson coefficients in SCET is determined
by the same virtual graphs as in the matching step. However, the evolution is not uniquely fixed given only 
the Wilson coefficients themselves; one needs also the mixing terms.
The Wilson coefficients satisfy a renormalization group equation of the 
form~\cite{Becher:2009qa,Gardi:2009qi,Dixon:2009ur,Catani:1998bh,Sterman:2002qn}
\begin{equation}
\label{Evolution_Hard}
   \frac{\rd}{\rd \ln \mu} \cC_{\bI}^\gGG (\mu) 
= 
    \GHIJ \cC_{\bJ}^\gGG (\mu)  \,,
\end{equation}
where, up to at least 2-loop order, $\GHIJ$ has the form
\begin{align}
  \label{Gamma_C_def}
\GHIJ(s,t,u,\mu) &=
    \left(\gcusp \frac{c_H}{2}
 \ln\frac{-t}{\mu^2} 
+ \gamma_H
- \frac{\beta(\alpha_s)}{\alpha_s}
 \right) \delta_{\bI \bJ} 
          + \gamma_{\mathrm{cusp}} \MIJ(s,t,u)  \,.
\end{align}
The $\beta(\alpha_s)/\alpha_s$ term compensates for the $g^2(\mu)$ dependence in the leading order Wilson coefficients.
The QCD beta function is $\beta(\alpha) = - 2 \alpha \left(\frac{\alpha}{4\pi} \right)\beta_0 + \cdots$
with $\beta_0$ given in Eq.~\eqref{b0def}.
The cusp anomalous dimension is $\gcusp =4\left(\frac{\alpha_s}{4\pi} \right) + \cdots$.
For NNLL resummation, one needs the  $\beta$-function and $\gcusp$ to 3-loop order, which
can be found in~\cite{Becher:2009th}, and the $\gamma_H$ and $\MIJ$ terms to 2-loop order, which we give below.

The specific form of the hard function RGE in Eq.~\eqref{Gamma_C_def} implies that
\begin{enumerate}
\item All of the $\mu$ dependence is proportional to the identity matrix in color space. It is fixed by a single Casimir $c_H$ which is known exactly. 
Explicitly,
\begin{eqnarray}
\label{c_H_def}
c_H =\sum_{i} C_{R_i}  = n_q C_F + n_g C_A    \,,
\end{eqnarray}
where $C_{R_i}$ is the quadratic Casimir of representation $R_i$, and $n_q( n_g)$ the total number
of quarks (gluons) in the hard process. The sum is over both initial and final states,
\item All of the color mixing terms are proportional to the universal cusp anomalous dimension $\gamma_{\mathrm{cusp}}$.  They 
are fixed by a single matrix $\MIJ(s,t,u)$ which depends on the channel and is known exactly.  
Once a color basis is chosen, this matrix can be computed from
\begin{eqnarray}
\mathbf{M} = 
  -\sum_{\langle {\red i}\ne {\red j} \rangle} \frac{\mathbf{T}_{\red i} \cdot \mathbf{T}_{\red j}}{2}\Big[ L(s_{\red ij})-L(t)\Big]  \,,
\end{eqnarray}
where $s_{\red 12}=s_{\red 34}=s$, $s_{\red 13}=s_{\red 24} = t$ and $s_{\red 24}=s_{\red 13}=u$ and $L(x)$ is given in Eq.~\eqref{Ldef}.
The operator $\mathbf{T}_{\red i} \cdot \mathbf{T}_{\red j}$ acts in color space on the operator basis,
as discussed in~\cite{Chiu:2009mg}.

\item $\gamma_H$ depends on the number of quarks and gluons involved, but not on the color structure.
It is given by
\begin{eqnarray}
\label{gH_def}
\gamma_H = n_q \gamma_q + n_g \gamma_g \, ,
\end{eqnarray}
where,
$\gamma_q = \left(\frac{\alpha_s}{4\pi}\right)( - 3 C_F)+\cdots$ and
$\gamma_g = \left(\frac{\alpha_s}{4\pi}\right)(-\beta_0)+\cdots$.
 These  quark and gluon anomalous dimensions can be found up to 3-loops in~\cite{Becher:2009qa}.
\end{enumerate}
The form of Eq.~\eqref{Gamma_C_def} is known to hold
up to 2-loop order. 
 Here, we verify it to 1-loop by direct calculation. The 2-loop results
are then used to solve the RGE of the Wilson coefficients for NNLL resummation.
Eq.~\eqref{Gamma_C_def} may hold to more than 2-loops, however at 3-loop order new possible terms may
arise which appear not to be forbidden by general arguments~\cite{Dixon:2009ur}. 

The universality of the off-diagonal terms in $\MIJ$ allows the mixing to be diagonalized once
and for all. This diagonalization was shown at next-to-leading order in the traditional approach
in~\cite{Kidonakis:1998nf}, and we find an eigensystem in SCET which is essentially identical.
Differences in the eigenvalues are due to different conventions for the diagonal terms, related to
the way double-counting of soft-collinear divergences in the two theories are handled.
Also, the explicit factor of $\log\frac{-t}{\mu^2}$ in  Eq.~\eqref{Gamma_C_def} 
is a convention. We could have used any other scale in this logarithm,
 such as $\log\frac{s}{\mu^2}$; the physical prediction is independent of our convention.

Let us denote the linear combination of Wilson coefficients which are
eigenvectors of $\MIJ$ by $\what{\cC}_{\bK}^{\gGG}$, with eigenvalues $\lambda_\bK$.
In the diagonal basis, the evolution equation takes the form
\begin{equation} \label{hard_rge}
  \frac{\rd}{\rd \ln \mu} \what{\cC}^{\gGG}_\bK(\mu) = 
    \left[\gamma_{\mathrm{cusp}} \frac{c_H}{2}\ln \frac{-t}{\mu^2}  
          + \gamma_H+ \gamma_{\mathrm{cusp}} \lambda_{\bK}
-\frac{\beta(\alpha_s)}{\alpha_s}
    \right]  \what{\cC}_{\bK}^{\gGG} (\mu)\, .
\end{equation}
As we can see, the evolution is now local in color space. Since the off-diagonal mixing terms are proportional
to $\gcusp$, this holds to at least order $\alpha_s^2$. Thus, it can be solved
in closed form once-and-for-all~\cite{Bauer:2003pi, Chiu:2009ft}.
 The solution to the RGE is
\begin{equation} \label{genH}
  \cC^{\gG}_{\bK}(\mu) = \frac{\alpha_s(\mu_h)}{\alpha_s(\mu)}\exp 
     \left[ c_H S (\mu_h, \mu)
            - A_H (\mu_h, \mu) 
            - A_{\Gamma} (\mu_h, \mu) 
\left( \lambda_\bK +  \frac{c_H}{2} \ln  \frac{- t}{\mu_h^2} \right) 
 \right]
\cC^{\gG}_{\bK}(\mu_h)   \,,
\end{equation}
where the functions $S (\nu, \mu)$ and $A (\nu, \mu)$ are the same as in~\cite{Becher:2009th}.
They have the general form
\begin{align}
S (\nu, \mu) &= - \int_{\alpha_s(\nu)}^{\alpha_s (\mu)} \rd \alpha
\frac{\gamma_{\mathrm{cusp}} (\alpha)}{\beta (\alpha)} \int_{\alpha_s
(\nu)}^{\alpha} \frac{\rd \alpha'}{\beta (\alpha')}\,\\
A_{\Gamma} (\nu, \mu) &= - \int_{\alpha_s (\nu)}^{\alpha_s (\mu)} \rd \alpha
\frac{\gamma_{\mathrm{cusp}} (\alpha)}{\beta (\alpha)}\,. \label{Adef}
\end{align}
$A_H (\nu, \mu)$ is the same  as $A_{\Gamma}(\nu,\mu)$ but with $\gamma_H$ replacing
$\gcusp$.
Closed-form expressions for these functions in renormalization-improved perturbation theory can be found in~\cite{Becher:2006mr}.

Since the RGE is independent of spin, we can also solve directly for the evolution of the spin-summed hard functions
$H_{\bI \bJ} = \sum_{\gG} \cC_{\bI}^\gG \cC_{\bJ}^{\gG \star}$.
If we go to the basis in which $\MIJ$ is diagonal, then the hard functions evolution equation
is particularly simple
\begin{equation}  
  \frac{\rd}{\rd \ln \mu} H_{\bK \bK'}(\mu) = 
    \left[\gcusp\left( c_H \ln \la\frac{t}{\mu^2}\ra  + \lambda_{\bK}+ \lambda_{\bK'}^\star \right)
          +2 \gamma_H -  \frac{2\beta(\alpha_s)}{\alpha_s}
    \right]  H_{\bK \bK'}(\mu)\, ,
\end{equation}
where we have used $\gamma_H^\star = \gamma_H$.
The solution, reinstating the $stu$ dependence for clarity, is
\begin{multline}
  H_{\bK \bK'}(s,t,u,\mu) = \frac{\alpha_s(\mu_h)^2}{\alpha_s(\mu)^2}\exp 
     \Big[ 2c_H S (\mu_h, \mu)
            -2 A_H (\mu_h, \mu) \Big] \\
\times \exp\left[
            -A_{\Gamma} (\mu_h, \mu) 
              \left(\lambda_{\bK}(s,t,u)+ \lambda_{\bK'}^\star(s,t,u) +   c_H \ln\la \frac{t}{\mu_h^2}\ra \right) 
     \right] H_{\bK \bK'}(s,t,u,\mu_h)   \,.
\end{multline}
Note that the hard function is Hermetian, by definition, but not real. Its Hermeticity is preserved by the renormalization group
evolution.

In the remainder of this section, we give the explicit expressions for $c_H$, $\gamma_H$, $\MIJ$, and $\lambda_\bK$ for the various channels.

\subsection{$qq \to qq$ channels \label{sec:RGEqq}}
The values of $c_H$ and $\gamma_H$ for all of the $qq\to qq$ crossings are
\begin{align}
 c_H &= 4 C_F\\
\gamma_H &= \left( \frac{\alpha_s}{4 \pi}\right) \left( - 12 C_F \right) + \cdots   \,,
\end{align}
which agree with Eqs.~\eqref{c_H_def} and \eqref{gH_def}.
The mixing matrix for the representative channel $qq^\prime \to qq^\prime$ is 
\begin{align}
  \MIJ(s,t,u) &= 
  \begin{pmatrix}
       4C_F[L(u) -L(s)]-C_A[L(t)+L(u)-2L(s)]  &    \quad 2[L(u)-L(s)]\\
       \frac{C_F}{C_A}[L(u)-L(s)] & 0 
  \end{pmatrix}  \,,
\end{align}
and the mixing matrix for other channels can be obtained by using the crossing relations given in 
Table~\ref{tab:crossqqqq}.

For conciseness, since the matrix does not depend on $\mu$, we will use the abbreviated notation,
following~\cite{Kidonakis:1998nf},
\begin{align}
  T &\equiv L(t)-L(s) \\
  U &\equiv L(u)-L(s)   \,.
\end{align}
Then
\begin{align}
  \MIJ(s,t,u) &= 
  \begin{pmatrix}
      4C_F U -C_A(T+U) &    \quad 2U\\
        \frac{C_F}{C_A}U & 0
 \end{pmatrix}  \,.
\end{align}

When identical particles are involved, one has to be careful to use 
the RGE appropriate for that
channel. Recall that we used a Fierz identity to 
rewrite the $\cO^{sut}$ operators as linear combinations of the $\cO^{stu}$ operators in Eq.~\eqref{Bmix}.
The operators $\cO^{sut}$ and  $\cO^{stu}$ evolve differently and to ensure this linear combination is 
not altered by evolution, we must have 
\begin{equation}
  {\mathbf M}(s,t,u) ={\mathbf B}\cdot {\mathbf M}(s,u,t)\cdot {\mathbf B}  \,,
\end{equation}
which is easily seen to hold. In fact, we have already used this relation to 
write all of the contributions to the $qq\to qq$ channel
in terms of a basis which evolves with $M_{\bI \bJ}(s,u,t)$, 
thus letting us calculate a single hard function for the $qq \to qq$ channel.

The eigenvalues of $\MIJ$ are
\begin{align}
  \lambda_{\bpm} &= \frac{C_A}{2}(U-T)- \frac{1}{C_A} U \bpm 
\sqrt{U T + \frac{1}{4}C_A^2 (T-U)^2}
  \,.
\end{align}
Here, $\bpm$ is shorthand for the two complex roots of the discriminant.
The diagonal basis is related to the original basis by
\begin{align} \label{Oqqdiag}
  \cC_{\bpm}^{\gG} &= 
\lambda_\bpm  \cC_{\bO}^\gG +\frac{C_F}{C_A}U\,\cC_{\bT}^\gG \, .
\end{align}
The normalizations are arbitrary.
To change to the diagonal basis for the Wilson coefficients or the hard functions, it is convenient
to use matrix notation
\begin{align}
  \cC_{\bpm}^\gG &= F_{\bpm \bI} \, \cC_\bI^\gG = (F \cdot \cC^\gG)_{\bpm}  \\
  H_{\bpm \bpm'} &= F_{\bpm \bI}\, F^{\star}_{\bpm' \bJ}\, H_{\bI \bJ}=(F\cdot H \cdot F^{\dagger})_{\bpm \bpm'} \, ,
\end{align}
with
\begin{align}
F_{\bpm \bI}(s,t,u) &=
\begin{pmatrix}
  \lambda_{\blue +} &\quad \frac{C_F}{C_A} U \\
  \lambda_{\blue -} &\quad \frac{C_F}{C_A} U \\
\end{pmatrix} \, .
\end{align}
The same matrix applies to all channels, with the crossing given in Table~\ref{tab:crossqqqq}. Note that
the explicit factor of $\ln\frac{- t}{\mu^2}=L(t)$ in Eq.~\eqref{Gamma_C_def} must be crossed as well. If $t$ crosses
into $s$, this logarithm has an imaginary part given by $L(s)$ in Eq.~\eqref{Ldef}. However, this particular imaginary 
part drops out when $\cC_{{\blue +}}^\gG$ and $\cC_{{\blue -}}^\gG$ are combined into the hard function so it can generally be 
ignored for physical applications. The imaginary parts in $F_{\bpm \bI}$ and $\lambda_\bpm$ are important and must be included.

\subsection{$gg \to q\bar{q}$ channels}
For the $gg \to q\bar{q}$  channels,
\begin{align}
 c_H &=  2C_F+2C_A\\
\gamma_H &= \left( \frac{\alpha_s}{4 \pi}\right) \left( - 6 C_F-2\beta_0 \right) + \cdots \, ,
\end{align}
and
\begin{align}\label{qgMIJ}
  \MIJ(s,t,u) &= 
  \begin{pmatrix}
      -C_F\,T         &\quad  0              &\quad  2(T- U)                \\
      0              &\quad  -(C_F + C_A)T + C_A U           &\quad   2 (U-T)\\
      -\frac{1}{2} U  &\quad -\frac{1}{2} T  &\quad  -(C_F + C_A)T \\
  \end{pmatrix} \,.
\end{align}
The eigenvalues are
\begin{align}
  \lambda_\bK = -(C_F+C_A)T + \wh\lambda_\bK \,,
\end{align}
where $\wh\lambda_\bK$ are  solutions of 
\begin{eqnarray}
\label{cubic}
\wh\lambda^3_\bK -C_A(T+U)\wh\lambda^2_\bK +\left[ C_A^2 T U- (T-U)^2\right] \wh\lambda_\bK + C_A(T-U)^2(T+U) =0 \,.
\end{eqnarray}
It is straightforward to solve this this cubic equation, with general complex coefficients, but the explicit solutions
are unilluminating. The Wilson coefficients and hard function in the diagonal basis are
\begin{align}
  \cC^{\gG}_\bK &= F_{\bK \bI} \cC^{\gG}_\bI \qquad
H_{\bK\bK'} 
=(F \cdot H \cdot F^{\dagger})_{\bK \bK'}  \, ,
\end{align}
with
\begin{align}
F_{\bK \bI}(s,t,u) &=
\begin{pmatrix}
(U-T) T - C_A U \wh\lambda_\bO + \wh\lambda_\bO^2 & \quad U(T-U) & \quad \frac{1}{2} U (C_A U - \wh\lambda_\bO) \\
(U-T) T - C_A U  \wh\lambda_\bT + \wh\lambda_\bT^2 & \quad U(T-U) & \quad \frac{1}{2} U (C_A U - \wh\lambda_\bT) \\
(U-T) T - C_A U \wh\lambda_\bTH + \wh\lambda_\bTH^2 & \quad U(T-U) & \quad \frac{1}{2} U (C_A U - \wh\lambda_\bTH) \\
  \end{pmatrix} \,.
\end{align}

\subsection{$gg\to gg$ channel}
\newcommand{\uca}{U}
\newcommand{\tca}{T}
\newcommand{\utca}{U-T}
\newcommand{\tuca}{T-U}
For $gg\to gg$,
\begin{align}
 c_H &=  4C_A\\
\gamma_H &= \left( \frac{\alpha_s}{4 \pi}\right) \left( -4\beta_0 \right)
+ \cdots \, .
\end{align}
The mixing matrix is given by Kidonakis et al. in~\cite{Kidonakis:1998nf}. Our color basis agrees with theirs, but
they have $t \leftrightarrow u$ compared to the convention of~\cite{Ellis:1985er} which we also use.
In addition, we use $\log\frac{-t}{\mu^2}$ rather than $\log\frac{s}{\mu^2}$ for the $\mu$-dependence piece,
which changes the diagonal elements of the matrix. In our notation, the mixing matrix is
\begin{align}
&~~~~~~~~~~~~~~~~~~~~~~~~~~~~~~~~~~~~~~~~~  \MIJ(s,t,u) = \\
&  \begin{pmatrix}
    C_A( U-2T)   & 0     & 0    & 0     & 0       & 0        & -\tca     & \utca   & 0  \\
    0   & -C_A T     & 0    & 0     & 0       & 0        & 0         & \tuca   & -\uca  \\
    0   & 0     & C_A (U-2T)    & 0     & 0       & 0        & -\tca     & \utca   & 0  \\
    0   & 0     & 0    & C_A(U-T)   & 0       & 0        & \tca      & 0       & \uca  \\
    0   & 0     & 0    & 0     & -C_A T       & 0        & 0         & \tuca   & -\uca  \\
    0   & 0     & 0    & 0     & 0       & C_A(U-T)      & \tca      & 0       & \uca  \\
  \utca & 0     &\utca & \uca  & 0       & \uca     & 2C_A (U-T)       & 0       & 0  \\
 -\tca  & -\uca &-\tca &  0    & -\uca   & 0        & 0         & -2C_A T       & 0  \\
    0   & \tuca &  0    & \tca  & \tuca  & \tca     & 0         &0        & 0 \\
  \end{pmatrix}\,. \nn
\end{align}
The eigenvalues and eigenvectors of this matrix are given for general $N$ in~\cite{Kidonakis:1998nf}. For $N=3$,
the results simplify, and we find
\begin{align}
  \lambda_{\blue 1} &=   \lambda_{\blue 2} = -3 T \\
  \lambda_{\blue 3} &=   \lambda_{\blue 4} = 3 (U-T)\\
  \lambda_{\blue 5} &=   \lambda_{\blue 6} = 3 (U-2T)\\
  \lambda_{\blue 7} &= 2(U-2T) \\
  \lambda_{{\blue 8},{\blue 9}} &=  2U-4T  \bpm 4 \sqrt{T^2- T U+U^2} \,.
\end{align}
The matrix of eigenvectors is
\begin{align}
&~~~~~~~~~~~~~~~~~~~~~~~~~~~~~~~~~~~~~  F_{\bK \bI}(s,t,u) = \\
&\begin{pmatrix}
   2U     & 
   7T     & 
   2U     &
   2(T-U) &
   7T     &
   2(T-U) &
   0      &
   -6U     &
   6(U-T) 
\\
  0       &
  1       &
  0       &
  0       &
  -1      &
  0       &
  0       &
  0       &
  0       
\\
  2U      &
  -2T     &
  2U      &
  7(U-T)  &
  -2T     &
  7(U-T)  &
  -6U     &
  0       &
  6T      &
\\
  0       &
  0       &
  0       &
  1       &
  0       &
  -1      &
  0       &
  0       &
  0       &
\\
7 U       &
2 T       &
7 U       &
2(U-T)     &
2 T       &
2 (U-T)   &
6 (T - U) &
-6 T      &
0
\\
  1       &
  0       &
  -1      &
  0       &
  0       &
  0       &
  0       &
  0       &
  0       &
\\
  1      &
  1      &
  1      &
  1      &
  1      &
  1      &
  -1       &
  -1       &
  -1       &
\\
a_1(\lambda_\bE) & a_2(\lambda_\bE) & a_3(\lambda_\bE) &
a_4(\lambda_\bE) & a_5(\lambda_\bE) & a_6(\lambda_\bE) &
a_7(\lambda_\bE) & a_8(\lambda_\bE) & a_9(\lambda_\bE) \\
a_1(\lambda_\bN) & a_2(\lambda_\bN) & a_3(\lambda_\bN) &
a_4(\lambda_\bN) & a_5(\lambda_\bN) & a_6(\lambda_\bN) &
a_7(\lambda_\bN) & a_8(\lambda_\bN) & a_9(\lambda_\bN) \\
  \end{pmatrix}\,,
\end{align}
with
\begin{align}
  a_1(\bl) &=
-\frac{4 T^2 U^2 (T-U)^2 \left(4 T^2-4 T U+5 U^2\right) \left(5 T^2-5 T U+8 U^2\right)}
{240 T^4+30 T^3 (\bl -16 U)+9 T^2 U (56 U-5   \bl )+T U^2 (35 \bl -264 U)+10 U^3 (6 U-\bl )}\nn \\
  a_2(\bl) &=
\frac{1}{3} T^2 (T-U) \left(T^2 (6 \bl -8 U)+T U (8 U-\bl )+5 U^2 (\bl -4 U)\right) \\
  a_3(\bl) &= a_1(\bl)\, \nn\\
  a_4(\bl) &=
\frac{1}{3} T (T-U)^2 \left(T^2 (8 U+6 \bl )-T U (8 U+11 \bl )+10 U^2 (2 U+\bl )\right) \nn\\
  a_5(\bl) &= a_2(\bl)\, \nn\\
  a_6(\bl) &= a_4(\bl)\, \nn\\
  a_7(\bl) &=
(T-U)^2 \left(-10 T^3 \bl +T^2 U (16 U+23 \bl )-T U^2 (26 U+23 \bl )+10 U^3 (2 U+\bl )\right) \nn \\
  a_8(\bl) &=
T^2 \left(-10 T^3 \bl +T^2 U (16 U+7 \bl )-T U^2 (6 U+7 \bl )+10 U^4\right) \nn \\
  a_9(\bl) &=
2 T^2 (T-U)^2 \left(4 T^2-4 T U+5 U^2\right) \nn \,.
\end{align}

\section{Soft function \label{sec:soft}}
The results for the NLO Wilson coefficients and their RG evolution apply for any physical process for
which massless $2\to 2$ scattering gives the leading order contribution. If we are interested in a particular
dijet observable, we must in addition calculate a process-dependent soft function. In general, soft function
calculations may be much more difficult than the hard function calculation, since they can involve arbitrarily
complicated phase space cuts. However, as we will show in this section, there are some universal features
which must hold for {\it any} soft function. In particular, the $\mu$-dependence of the soft function
and its RGE to 2-loop order are almost completely fixed.

In SCET, a soft function is calculated by taking matrix elements of the products of
Wilson lines which appear in the SCET operators. For example, for the 
$q q'\to q q'$ operators, these are
\begin{eqnarray}
 \mathcal{W}_{\bO} &=& \tmmathbf{T} \left\{ (Y^\dagger_\rF {\blue \tau^a} Y_\rT )(   Y^\dagger_\rTH {\blue \tau^a} Y_\rO)  \right\}
 \nn\\
 \mathcal{W}_{\bT} &=& \tmmathbf{T} \left\{ (Y^\dagger_\rF {\blue \mathbf{1}} Y_\rT ) (Y^\dagger_\rTH{\blue \mathbf{1}} Y_\rO)  \right\}\,.
\end{eqnarray}
The soft function is then calculated from
\begin{equation}
    S_{\bI \bJ} \left( \{k\}, {\red n^{\mu}_i},\mu \right) 
= 
    \sum_{X_s}
    \bra{0   } \mathcal{W}_\bI^\dagger \ket{X_s}
    \bra{X_s } \mathcal{W}_\bJ         \ket{0}  
    F_S (\{k\})  \,,
\end{equation}
where the sum is over soft radiation in the final state. The function $F_S(\{k\})$ 
encodes  the dependence on various projections on the soft momenta related
to the definition of the observable. No matter what the observable
is, the soft function can only depend on directions ${\red n_{i^{}}^{\mu}}$ 
of the various Wilson lines, and on arbitrary soft scales $\{ k\}$
relevant to the projections.
 Because of factorization, it cannot depend on the
energy of the jets, the hard scales $s, t, u$ or the energy fractions $x_i$ of
the PDFs. 

For the final cross section to be independent of $\mu$,  the soft function must
satisfy a renormalization group equation closely related to that of the hard Wilson
coefficients. The soft function evolution will not be local in $\{k\}$, but
must be local in Laplace space for factorization to hold. 
The Laplace transform of the soft function is defined, for one scale $k$, as
\begin{align}
   \widetilde{S}_{\bI \bJ}(Q, {\red n^{\mu}_i},\mu) 
   = 
   \int_0^\infty \rd k \exp\left(-\frac{k}{Q e^{\gamma_E}}\right) 
   S_{\bI \bJ}(k, {\red n^{\mu}_i},\mu)\,,
\end{align}g
with the natural generalization to more scales $\{ k \}$.
In Laplace space, the jet function and PDF evolution is also local.
Furthermore, since
the jet functions and the PDFs evolve diagonally in color, the color mixing terms
in the soft function evolution must exactly compensate the color mixing terms
in the hard function evolution.

It follows that the soft function must satisfy
\begin{equation}
 \frac{\rd}{\rd \ln \mu} \widetilde{S}_{\bI \bJ} \Big(\{ Q \}, {\red n^{\mu}_i},\mu\Big) = 
  -\widetilde{S}_{\bI \bL} \Big(\{ Q \}, {\red n_i^\mu}, \mu\Big) \Gamma^S_{\bL \bJ}
  -\Gamma^{S\dagger}_{\bI \bL}\widetilde{S}^\dagger_{\bL \bJ} \Big(\{ Q \},{\red n^{\mu}_i}, \mu\Big)  \,,
\end{equation}
where
\begin{equation}
 \Gamma^S_{\bI \bJ} = 
          \left( \gcusp c_Q \ln  \frac{\{Q\}} {\mu} + \gcusp\,r( {\red n_i^\mu} )
+ \gamma_S \right) \delta_{\bI \bJ} 
          + \gcusp \MIJ ({\red n_i^\mu}) \, .
\end{equation}
This has many of the same properties as the hard function RGE, Eq.~\eqref{Gamma_C_def}:
\begin{enumerate}
\item The $\mu$-dependence is proportional to $\gamma_{\mathrm{cusp}}$ times the identity matrix in color space.
\item The color mixing is proportional to  $\gamma_{\mathrm{cusp}}$ and fixed by a single matrix.
\item The unknown quantity $\gamma_S$ depends only on the channel, not the color structure.
\end{enumerate}
Furthermore, for the soft function we can observe further that 
\begin{enumerate}
\setcounter{enumi}{3}
\item The matrix $\MIJ ( {\red n_i} )$ can be taken 
to be the same as the $\MIJ$ matrix appearing in Eq.~\eqref{Gamma_C_def}, which depends on $s,t$ and $u$.
This is possible because only dimensionless Lorentz-invariant ratios
appear, and $({\red n_i} \cdt {\red n_j}) / ({\red n_k} \cdt {\red n_l})=( p_i \cdt p_j)/ (p_k \cdt p_l)$.
To make this transparent, we can even write
%
\begin{eqnarray}
\mathbf{M} = 
  -\sum_{\langle {\red i}\ne {\red j} \rangle} \frac{\mathbf{T}_{\red i} \cdt \mathbf{T}_{\red j}}{2}
\Big[ L(n_{\red {ij}}) - L(n_{\red {13}}) \Big]
\end{eqnarray}
where $n_{\red{ij}}= \pm {\red n_i} \cdt {\red n_j}$, taking the $+$ sign when
${\red n_i}$ and ${\red n_j}$ are
both incoming or both outgoing, and the $-$ sign otherwise.
This form of $\mathbf{M}$ is a variation of the form presented in \cite{Chiu:2009mg}.
\item  The color-diagonal piece, proportional to $\gamma_{\mathrm{cusp}} \delta_{\bI \bJ}$, may have dependence on the 
directions ${\red n_i^\mu}$ and on the soft scales $\{Q\}$.
\item RG invariance lets us solve for $\gamma_S$ in terms of $\gamma_H$ and the jet and
PDF anomalous dimensions $\gamma_J$ and $\gamma_f$. 
\end{enumerate}

Note that in defining the matrix $\MIJ$ with the hard function, Eq.~\eqref{Gamma_C_def},
there was an implicit convention in the diagonal terms. For example, if we had written $\ln\frac{-u}{\mu^2}$ instead of
$\ln\frac{-t}{\mu^2}$ the diagonal terms in $\MIJ$ would be different. This change of convention
can then be absorbed in the $r( {\red n_i^\mu} )$ term in the soft function anomalous dimension. The terms
in the soft function evolution equation proportional to $\delta_{\bI \bJ}$ are also different from those in the
hard function evolution equation because they must additionally
 compensate for the evolution of the jet functions
and PDFs. Since the number and type of jets which contribute depends on the observable, this term is observable-dependent.
Once an observable is defined, the entire soft function RGE can be fixed either by a 1-loop
computation or by analyzing the factorization formula. 

Since the mixing terms in the soft function are proportional to $\MIJ$, there is no mixing in the basis in which
$\MIJ$ is diagonal. In the previous section, explicit expressions were given for 
the matrices $F_{\bI \bK}$ which let us change from the $\bI$ basis, in which the NLO matching
 is simple. to the $\bK$ basis, in which the evolution is simple.
We had written
\begin{align}
  H_{\bK \bK'} = (F \cdot H \cdot F^{\dagger})_{\bK \bK'} \,.
\end{align}
Consequently the soft function transforms as
\begin{align}
  S_{\bK \bK'} = [(F^{-1})^{\dagger} \cdot S \cdot F^{-1}]_{\bK \bK'}\,.
\end{align}
The cross section involves a sum over color structures
\begin{align}
\sum_{\bI \bJ} H_{\bI \bJ} S_{\bJ \bI}  = \sum_{\bK \bK'} H_{\bK \bK'} S_{\bK' \bK}= \text{Tr} (H S) \, ,
\end{align}
which is basis independent. Since both $H$ and $S$ are Hermetian, this sum is real. However, note that
$F$ may not be unitary since $\MIJ$ is in general not Hermetian.

The general solution to the soft function RGE in the diagonal basis is
\begin{multline} \label{genS}
  \widetilde{S}_{\bK\bK'}(\{Q\},{\red n_i^\mu},\mu) = \exp 
     \Big[ -2c_Q  S(\mu_s, \mu)  +2 A_S (\mu_s, \mu)\Big]\\
\times 
\exp \Big[   A_{\Gamma} (\mu_s, \mu)  
          \left(   
             \lambda_\bK 
            +\lambda_{\bK'}^\star  
            + r( {\red n_i^\mu} ) 
            + r( {\red n_i^\mu} )^\star 
            + 2c_Q \ln  \frac{ \{Q\}}{\mu_s} 
          \right)
     \Big] 
\widetilde{S}_{\bK\bK'}(\{Q\},{\red n_i^\mu},\mu_s) \,.
\end{multline}
This can be transformed back to momentum space using the techniques described,
for example, in~\cite{Becher:2008cf}, but the general expression is not particularly
illuminating.
Note that even in the $\bK$ basis, in which the evolution is diagonal, the soft function
itself may still have non-diagonal terms ($\bK \neq \bK'$) thus leading to color mixing 
among the Wilson coefficients $\cC_{\bK}^{\gG}$.
For next-to-leading-log resummation (NLL), the
soft function is completely determined by its anomalous dimensions, and then
$\widetilde{S}_{\bK\bK'}(\{Q\},\mu)$ is diagonal for any $\mu$. However,
at NNLL the additional non-diagonal terms in the finite parts of the soft function have to
be included. Nevertheless, due to the very general arguments presented above,
each component of
the soft function at the scale $\mu$ is proportional to the the same
component at any other scale $\mu'$,  $\widetilde{S}_{\bK \bK'}(\mu)\propto \widetilde{S}_{\bK \bK'}(\mu')$,
as this equation shows.

Although a factorization formula will let us deduce $\gamma_S$ from RG invariance, 
it is satisfying to also calculate this anomalous dimension directly.
Such a calculation provides an important check that the observable is well defined
and that the factorization theorem actually works.
Moreover, performing the calculation of a particular
soft function to order $\alpha_s$ will also produce the $\mu$-independent $\alpha_s$ terms,
which are necessary for NNLL resummation. An example can be found in~\cite{Kelley:wip}.
Nevertheless, it is easy to imagine a complicated enough
observable for which direct calculation is not possible. In that case, these general results will
allow us to resum the next-to-leading logarithms (NLL) without actually doing the soft integrals.

\section{Conclusions \label{sec:conc}}
We have presented the tools necessary for an effective field theory calculation at NNLL 
of observables based on $2\to2$ processes in QCD.  The 1-loop matching coefficients for all the $2 \to 2$ 
processes  have been compiled in a format
that is conducive to the computation of dijet observables in SCET. 
The evolution of these
matching coefficients is independent of the observable and we have given analytic formulas for the
Wilson coefficients at an arbitrary scale $\mu$ including NNLL resummation.

For each channel, there is a basis in which the evolution of the Wilson coefficients is diagonal in color space.
These bases and the associated eigenvalues of the evolution equations were given explicitly for
all channels.
Unfortunately, the basis in which the 1-loop matching is simple and the basis in which the renormalization
group evolution is simple do not coincide. Nevertheless, since we have provided the explicit matrices
to go between bases, numerical evaluation of the resummed expressions is now straightforward.

General soft functions, which may depend on the observable, were also examined. They were shown to have 
similar universal properties to the hard function as required by the consistency of the effective field theory.
The bases which diagonalize the Wilson coefficient evolution also diagonalize the soft functions
evolution and simplify the form of the resummed expressions.

With these results, it will be possible to begin accurate calculations of dijet observables
including NNLL resummation.  As a start, it would be useful to compute a
simple observable which can check the RG evolution for the soft function we have derived here
through explicit calculation. Such a calculation is given in~\cite{Kelley:wip}. Then, 
more interesting observables may be approached, including, for example,
the jet $p_T$ spectrum at large $p_T$, jet substructure,
such as jet masses or angularities~\cite{Ellis:2010rw}, hadronic event shapes~\cite{Banfi:2004yd}, or even possibly 
observables related to the color flow in an event~\cite{Gallicchio:2010sw}.

\section{Acknowledgements}
We would like to thank Thomas Becher, Andreas Fuhrer, Lance Dixon and Aneesh Manohar for helpful discussions.
Our research is supported in part by the Department of Energy, under grant DE-SC003916.


\begin{thebibliography}{99}


\bibitem{Hornig:2009vb}
  A.~Hornig, C.~Lee and G.~Ovanesyan,
  JHEP {\bf 0905}, 122 (2009)
  [arXiv:0901.3780 [hep-ph]].

\bibitem{Schwartz:2007ib}
  M.~D.~Schwartz,
  Phys.\ Rev.\  D {\bf 77}, 014026 (2008)
  [arXiv:0709.2709 [hep-ph]].

\bibitem{Manohar:2003vb}
  A.~V.~Manohar,
  Phys.\ Rev.\  D {\bf 68}, 114019 (2003)
  [arXiv:hep-ph/0309176].

\bibitem{Idilbi:2005ky}
  A.~Idilbi and X.~d.~Ji,
  Phys.\ Rev.\  D {\bf 72}, 054016 (2005)
  [arXiv:hep-ph/0501006].

\bibitem{Becher:2009th}
  T.~Becher and M.~D.~Schwartz,
  JHEP {\bf 1002}, 040 (2010)
  [arXiv:0911.0681 [hep-ph]].

\bibitem{Ahrens:2010zv}
  V.~Ahrens, A.~Ferroglia, M.~Neubert, B.~D.~Pecjak and L.~L.~Yang,
  arXiv:1003.5827 [hep-ph].

\bibitem{Idilbi:2005ni}
  A.~Idilbi, X.~d.~Ji, J.~P.~Ma and F.~Yuan,
  Phys.\ Rev.\  D {\bf 73}, 077501 (2006)
  [arXiv:hep-ph/0509294].

\bibitem{Idilbi:2006dg}
  A.~Idilbi, X.~d.~Ji and F.~Yuan,
  Nucl.\ Phys.\  B {\bf 753}, 42 (2006)
  [arXiv:hep-ph/0605068].



\bibitem{Becher:2007ty}
  T.~Becher, M.~Neubert and G.~Xu,
  JHEP {\bf 0807}, 030 (2008)
  [arXiv:0710.0680 [hep-ph]].

\bibitem{Becher:2008cf}
  T.~Becher and M.~D.~Schwartz,
  JHEP {\bf 0807}, 034 (2008)
  [arXiv:0803.0342 [hep-ph]].

\bibitem{Ahrens:2008nc}
  V.~Ahrens, T.~Becher, M.~Neubert and L.~L.~Yang,
  Eur.\ Phys.\ J.\  C {\bf 62}, 333 (2009)
  [arXiv:0809.4283 [hep-ph]].

\bibitem{Chien:2010kc}
  Y.~T.~Chien and M.~D.~Schwartz,
  arXiv:1005.1644 [hep-ph].

\bibitem{Abbate:2010xh}
  R.~Abbate, M.~Fickinger, A.~H.~Hoang, V.~Mateu and I.~W.~Stewart,
  arXiv:1006.3080 [hep-ph].

\bibitem{Kaplan:2008pt}
  D.~E.~Kaplan and M.~D.~Schwartz,
  Phys.\ Rev.\ Lett.\  {\bf 101}, 022002 (2008)
  [arXiv:0804.2477 [hep-ph]].

\bibitem{Chiu:2008vv}
  J.~y.~Chiu, R.~Kelley and A.~V.~Manohar,
  Phys.\ Rev.\  D {\bf 78}, 073006 (2008)
  [arXiv:0806.1240 [hep-ph]].

\bibitem{Fuhrer:2010eu}
  A.~Fuhrer, A.~V.~Manohar, J.~y.~Chiu and R.~Kelley,
  arXiv:1003.0025 [hep-ph].

\bibitem{Kunszt:1993sd}
  Z.~Kunszt, A.~Signer and Z.~Trocsanyi,
  Nucl.\ Phys.\  B {\bf 411}, 397 (1994)
  [arXiv:hep-ph/9305239].

\bibitem{Bern:1990cu}
  Z.~Bern and D.~A.~Kosower,
  Phys.\ Rev.\ Lett.\  {\bf 66}, 1669 (1991).

\bibitem{Kidonakis:1998nf}
  N.~Kidonakis, G.~Oderda and G.~F.~Sterman,
  Nucl.\ Phys.\  B {\bf 531}, 365 (1998)
  [arXiv:hep-ph/9803241].

\bibitem{Chiu:2009mg}
  J.~y.~Chiu, A.~Fuhrer, R.~Kelley and A.~V.~Manohar,
  Phys.\ Rev.\  D {\bf 80}, 094013 (2009)
  [arXiv:0909.0012 [hep-ph]].

\bibitem{Becher:2009cu}
  T.~Becher and M.~Neubert,
  Phys.\ Rev.\ Lett.\  {\bf 102}, 162001 (2009)
  [arXiv:0901.0722 [hep-ph]].

\bibitem{Becher:2009qa}
  T.~Becher and M.~Neubert,
  JHEP {\bf 0906}, 081 (2009)
  [arXiv:0903.1126 [hep-ph]].

\bibitem{Gardi:2009qi}
  E.~Gardi and L.~Magnea,
  JHEP {\bf 0903}, 079 (2009)
  [arXiv:0901.1091 [hep-ph]].

\bibitem{Dixon:2009ur}
  L.~J.~Dixon, E.~Gardi and L.~Magnea,
  JHEP {\bf 1002}, 081 (2010)
  [arXiv:0910.3653 [hep-ph]].

\bibitem{Dixon:2010hy}
  L.~J.~Dixon, E.~Gardi and L.~Magnea,
  arXiv:1001.4709 [hep-ph].


\bibitem{Catani:1998bh}
  S.~Catani,
  Phys.\ Lett.\  B {\bf 427}, 161 (1998)
  [arXiv:hep-ph/9802439].


\bibitem{Sterman:2002qn}
  G.~Sterman and M.~E.~Tejeda-Yeomans,
  Phys.\ Lett.\  B {\bf 552}, 48 (2003)
  [arXiv:hep-ph/0210130].


\bibitem{Ellis:1985er}
  R.~K.~Ellis and J.~C.~Sexton,
  Nucl.\ Phys.\  B {\bf 269}, 445 (1986).

\bibitem{Bauer:2000yr}
  C.~W.~Bauer, S.~Fleming, D.~Pirjol and I.~W.~Stewart,
  Phys.\ Rev.\  D {\bf 63}, 114020 (2001)
  [arXiv:hep-ph/0011336].

\bibitem{Bauer:2001yt}
  C.~W.~Bauer, D.~Pirjol and I.~W.~Stewart,
  Phys.\ Rev.\  D {\bf 65}, 054022 (2002)
  [arXiv:hep-ph/0109045].

\bibitem{Beneke:2002ph}
  M.~Beneke, A.~P.~Chapovsky, M.~Diehl and T.~Feldmann,
  Nucl.\ Phys.\  B {\bf 643}, 431 (2002)
  [arXiv:hep-ph/0206152].

\bibitem{Jouttenus:2009ns}
  T.~T.~Jouttenus,
  Phys.\ Rev.\  D {\bf 81}, 094017 (2010)
  [arXiv:0912.5509 [hep-ph]].

\bibitem{Cheung:2009sg}
  W.~Y.~Cheung, M.~Luke and S.~Zuberi,
  Phys.\ Rev.\  D {\bf 80}, 114021 (2009)
  [arXiv:0910.2479 [hep-ph]].

\bibitem{Ellis:2010rw}
  S.~D.~Ellis, C.~K.~Vermilion, J.~R.~Walsh, A.~Hornig and C.~Lee,
  arXiv:1001.0014 [hep-ph].

\bibitem{Bauer:2003pi}
  C.~W.~Bauer and A.~V.~Manohar,
  Phys.\ Rev.\  D {\bf 70}, 034024 (2004)
  [arXiv:hep-ph/0312109].
  
\bibitem{Chiu:2009ft}
  J.~y.~Chiu, A.~Fuhrer, R.~Kelley and A.~V.~Manohar,
  Phys.\ Rev.\  D {\bf 81}, 014023 (2010)
  [arXiv:0909.0947 [hep-ph]].

\bibitem{Becher:2006mr}
  T.~Becher, M.~Neubert and B.~D.~Pecjak,
  JHEP {\bf 0701}, 076 (2007)
  [arXiv:hep-ph/0607228].

\bibitem{Kelley:wip}
  R.~Kelley and M.~D.~Schwartz, {\it in preparation}.

\bibitem{Banfi:2004yd}
  A.~Banfi, G.~P.~Salam and G.~Zanderighi,
  JHEP {\bf 0503}, 073 (2005)
  [arXiv:hep-ph/0407286].

\bibitem{Gallicchio:2010sw}
  J.~Gallicchio and M.~D.~Schwartz,
  Phys.\ Rev.\ Lett.\  {\bf 105}, 022001 (2010)
  [arXiv:1001.5027 [hep-ph]].

\end{thebibliography}
\end{document}